\documentclass[ %
 reprint,
 superscriptaddress,
showpacs,preprintnumbers,
 amsmath,amssymb,
 aps,
 10pt,
 prb,
twocolumn
]{revtex4-1}

\usepackage{graphicx}
\usepackage{dcolumn,overpic}
\usepackage{bm}

\usepackage{hyperref}
\hypersetup{colorlinks=true,breaklinks,linkcolor=blue,urlcolor=blue,citecolor=blue} 

\usepackage[caption=false]{subfig}  
\usepackage{natbib}
\usepackage{mathtools}

\DeclarePairedDelimiter{\bra}{\langle}{\rvert}
\DeclarePairedDelimiter{\ket}{\lvert}{\rangle}

\begin{document}


\title{Analytic continuation by averaging Pad\'{e} approximants}

\author{Johan Sch\"ott}
\affiliation{Dept.\ of Physics and Astronomy, Uppsala University, Box 516, SE-75120 Uppsala, Sweden}
\author{Inka L. M. Locht}
\affiliation{Dept.\ of Physics and Astronomy, Uppsala University, Box 516, SE-75120 Uppsala, Sweden}
\affiliation{Institute of Molecules and Materials, Radboud University of Nijmegen, Heyendaalseweg 135, 6525 AJ Nijmegen, The Netherlands}
\author{Elin Lundin}%
\affiliation{Dept.\ of Physics and Astronomy, Uppsala University, Box 516, SE-75120 Uppsala, Sweden}
\author{Oscar Gr\r{a}n{\"a}s}%
\affiliation{Dept.\ of Physics and Astronomy, Uppsala University, Box 516, SE-75120 Uppsala, Sweden}
\affiliation{School of Engineering and Applied Sciences, Harvard University, Cambridge, Massachusetts 02138, USA}
\author{Olle Eriksson}
\affiliation{Dept.\ of Physics and Astronomy, Uppsala University, Box 516, SE-75120 Uppsala, Sweden}
\author{Igor Di Marco}
\affiliation{Dept.\ of Physics and Astronomy, Uppsala University, Box 516, SE-75120 Uppsala, Sweden}





\date{\today}

\begin{abstract}
The ill-posed analytic continuation problem for Green's functions and self-energies is investigated by revisiting the Pad\'{e} approximants technique. We propose to remedy to the well-known problems of the Pad\'{e} approximants by performing an average of several continuations, obtained by varying the number of fitted input points and Pad\'{e} coefficients independently. The suggested approach is then applied to several test cases, including Sm and Pr atomic self-energies, the Green's functions of the Hubbard model for a Bethe lattice and of the Haldane model for a nano-ribbon, as well as two special test functions. The sensitivity to numerical noise and the dependence on the precision of the numerical libraries are analysed in detail. The present approach is compared to a number of other techniques, i.e. the non-negative least-square method, the non-negative Tikhonov method and the maximum entropy method, and is shown to perform well for the chosen test cases. This conclusion holds even when the noise on the input data is increased to reach values typical for quantum Monte Carlo simulations. The ability of the algorithm to resolve fine structures is finally illustrated for two relevant test functions. 

\end{abstract}

\pacs{71.10.Fd, 71.15.Dx, 02.70.Hm, 05.30.Fk}
\maketitle


\section{\label{sec:intro}Introduction}
\begin{figure}[b]
\includegraphics[]{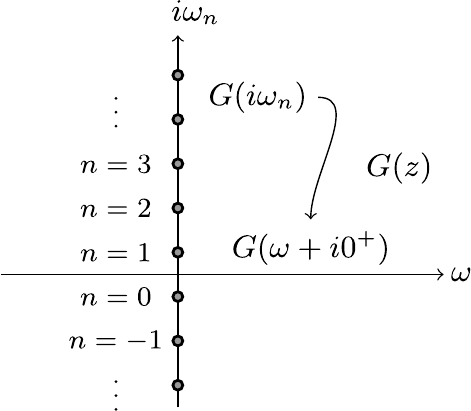}
\caption{Typical example of analytic continuation of the Green's function in the complex plane. The Matsubara frequencies on the imaginary axis are labelled following the convention explained in section~\ref{sec:green}.}
\label{figtikz}
\end{figure}

Strongly correlated materials are currently of great interest due to that they exhibit a plethora of exotic effects which may be important for technological applications\cite{highTc_ref,Jonker1950337}. One of the crucial problems behind the investigation of strongly correlated materials is determining their electronic structure. To this aim several computational methods have been developed in the last twenty years, including the dynamical mean-field theory (DMFT)~\cite{georges96rmp68:13,kotliar06rmp78:865} or the GW approach~\cite{aryasetiawan98rmp61:237}. These methods are often based on the Green's function formalism, and are not always applied directly to real energies. In fact, for technical reasons, it may be more convenient to work with complex energies and then obtain the Green's functions (and related observables) for real energies by means of analytic continuation, as illustrated in Fig. \ref{figtikz}.

The analytic continuation of functions whose formulas are known is a rather simple task in complex analysis. In numerical problems, instead, difficulties arise. Although there still exists a unique continuation if a function is known at infinitely many discrete points on the imaginary axis and with infinite precision~\cite{unique}, these conditions are not fulfilled in standard computations. The problem depicted in Fig. \ref{figtikz} is therefore ill-posed, especially if there is no a priori knowledge of the function structure on the real axis. Owing to these difficulties, several methods were proposed to perform the analytic continuation of Green's functions and self-energies. The most celebrated approaches include the maximum entropy method (MEM)\cite{PhysRevB.41.2380,Jarrell1996133,bryans,PhysRevB.44.6011,PhysRevE.81.056701,PhysRevE.82.026701,PhysRevB.81.155107,PhysRevB.82.165125}, the Singular Value Decomposition (SVD)\cite{PhysRevLett.75.517,PhysRevB.82.165125} and the closely related non-negative Tikhonov (NNT) method\cite{Tikhonov}, the non-negative least-square (NNLS) method \cite{NNLS}, the method of consistent constrains\cite{MCC}, stochastic regularization methods\cite{2006cond.mat.12233K} and sampling methods\cite{PhysRevB.76.035115,PhysRevB.57.10287,PhysRevB.62.6317,staar14}. Additionally, recent work shows the analytic continuation can be performed from complex-time, reducing its ill-conditioned property\cite{Little_wicks}. All these methods contributed greatly to advance our understanding of the analytic continuation, but in practical terms they still suffer of drawbacks such as requiring prior information on the function to continue, smearing high energy states, involving great computational efforts or applicability.  

Another well-known technique for analytic continuation is the Pad\'{e} approximant method. It consists in parametrising the numerical function by means of a ratio of two polynomials, or equivalently by a terminating continued fraction. There are different schemes for finding the Pad\'{e} approximant. One is Thiele's reciprocal difference method~\cite{Baker_with_Thiele}, which is a numerically fast method for attaining the function values at selected points in the complex plane by recursion. It was first applied to condensed matter physics by Vidberg and Serene\cite{vidberg77jltp29:179} to address the Eliashberg equations. A second approach is to determine the polynomial coefficients explicitly, giving the function everywhere in the complex plane. In the scheme proposed by Beach {\it{et al.}}~\cite{beach}, the Pad\'{e} coefficients are calculated through a standard matrix problem, which makes it possible to use widely available routines for linear algebra.

In general, all the schemes based on the Pad\'{e} approximant method have the advantage of not requiring any prior information on the function that one intends to continue. However, such an unbiased continuation often results in artefacts and spurious features. In the worst cases the continued function may violate important physical constraints, e.g. leading to a positive imaginary part of the Green's function. These problems are more severe for Pad\'{e} schemes based on recursive algorithms, due to the propagation of errors. Increasing the precision of the input data as well as that of the numerical routines employed in the algorithms removes spurious features from the continuation~\cite{beach}. Unfortunately, obtaining input data with a precision higher than double precision (about 64 bits) is not so straightforward for problems of current scientific interest. Conversely, one must often work with Green's functions and self-energies plagued by numerical noise, as arising from quantum Monte-Carlo (QMC) methods~\cite{gull11}. Hence, the Pad\'{e} approximant method has achieved less attention than the other techniques outlined above. 

The errors in the Pad\'{e} approximant method are due to the presence of approximate zero-pole pairs in the polynomials~\cite{beach}. These pairs arise from including too many points in the Pad\'{e} fitting and should ideally cancel each other when taking the ratio between polynomials. The limited precision prevents this from happening, leading to a sort of overfitting problem. Beach {\it{et al.}} proposed a procedure to identify the ideal number of points to include in the algorithm but this approach is hard to apply unless Matsubara data of very high precision is available~\cite{beach}. More recently, in Refs.~\onlinecite{Andreas,2011CoPhC.182..448S}, the authors suggested some heuristic procedures to identify and eliminate the zero-pole pairs from the polynomials. Despite representing an improvement with respect to the original formulation, all these schemes still suffer of a certain degree of arbitrariness and require an ad-hoc analysis, case by case. 

In the present work, we show that the overfitting problem can be greatly reduced by reformulating the Beach algorithm to use fewer Pad\'{e} coefficients than input points, which has been suggested already in Ref.~\onlinecite{PhysRevB.87.245135}. This implies transforming the problem into a least square (LS) minimisation. We build on this approach to tackle the sensitivity of the Pad\'{e} method by constructing a set of analytic continuations where the number of coefficients and input points are varied independently.  These continuations can then be used to perform a weighted average in which the spurious zero-pole pairs have a strongly reduced contribution. We show that this procedure stabilises the Pad\'{e} method and removes unphysical structures. The advantages of our approach are shown to be especially important for low input precision data. Even for numerical noise comparable to standard QMC simulations our 
Pad\'{e} scheme provides good analytic continuations. A further improvement is obtained by exploiting the symmetries of the Green's function, i.e. by considering a small number of negative Matsubara frequencies. Finally, we benchmark our approach on various test-cases, including the atomic self-energies of Sm and Pr, the Green's functions of the Hubbard model for a Bethe lattice and of the Haldane model for a nano-ribbon, as well as some simple test functions. These results are compared with exact solutions as well as with analytic continuations obtained via MEM, NNT and NNLS. The proposed Pad\'{e} method is shown to often perform better than the other methods over a wide range of noise magnitudes. 

The paper is organised as follows. Section I is an introduction, section~\ref{sec:green} is dedicated to a short review of the relevant properties of Green's functions and self-energies which we intend to continue. In section~\ref{sec:pade}, our Pad\'{e} scheme is described in detail. The results are illustrated in section~\ref{sec:result}, together with an description of step-by-step improvements, tests of the accuracy and stability with respect to Matsubara noise and precision of the numerical routines. A wider range of tests as well as a comparison with some other methods for analytic continuation are presented in section~\ref{sec:OtherAC}. Finally the resolving power of our algorithm is analysed in section~\ref{sec:power}, which is followed by the conclusions.

 \section{\label{sec:green}Green's function formalism}
The one-particle Green's function is a key quantity in the solution of models exhibiting strong correlations. Here we are going to briefly summarise the main properties of the Green's function, which will be then used in section \ref{sec:pade} for explaining and improving the Pad\'e approximant method.  In finite temperature Matsubara formalism, the one-electron Green's function of a time-independent Hamiltonian is written as 
\begin{equation}
G(\tau)=-\langle T[c(\tau)c^\dagger(0)] \rangle
\label{eq:gf_tau}
\end{equation}
where $\tau$ is the imaginary time, $T$ is the time ordering super-operator, $c(\tau)$ and $c(\tau)^\dagger$ are respectively annihilation and creation operators in Heisenberg representation. For simplicity we consider here models with a single orbital, so that no subscripts are needed. The generalisation to multi-orbital systems is straightforward. The expectation value in Eq.~\eqref{eq:gf_tau} denotes the thermal average over the grand canonical ensemble, and $G(\tau)$ is defined on the interval $\tau \in (-\beta,\beta]$, where $\beta$ is the inverse temperature. A fermionic Green's function is anti-periodic due to the anti-commutation relation $\left \{ c,c^\dagger \right \}=1$ and the trace invariance under cyclic permutations. By periodically repeating $G(\tau)$, its Fourier representation is
\begin{equation}
G(\tau) = \frac{1}{\beta} \sum_{n=-\infty}^{\infty} G_n e^{-i \omega_n \tau} 
\end{equation}
with Fourier coefficients 
\begin{equation}
G_n = \int_{0}^{\beta}   \text{d} \tau e^{i \omega_n \tau} G(\tau)  
\end{equation}
and fermionic Matsubara frequencies $\omega_n=(2n-1)\pi / \beta$.
A unique spectral function $\rho(\omega) \in \mathbb{R}^+$ exists for a complete set of $G_n$-values, related to $G_n$ through the Hilbert transform
\begin{equation}
G(z) = \int_{-\infty}^{\infty}   \text{d} \omega \frac{1}{z-\omega}\rho(\omega)
\label{eq:hilbert}
\end{equation}
by setting $z=i \omega_n$. Time reversal symmetry is ensured by $G(z^*)=G(z)^*$ and follows from Eq.~\eqref{eq:hilbert}, as is shown in Ref.~\onlinecite{beach}. Due to causality, the Green's function is analytic in the whole complex plane except along the real axis where its imaginary part has a discontinuity. The spectral function in Eq. \eqref{eq:hilbert} can be expressed as
\begin{equation}
\rho(\omega) = -\frac{1}{\pi} \text{Im}[G(\omega+i\delta^+)]  \quad \text{with} \quad \delta \rightarrow 0^+,
\label{eq:spec_delta}
\end{equation}
and is $1/(2\pi)$ times the magnitude of the discontinuity in $\text{Im}[G(z)]$ on the real axis. The parameter $\delta$ in Eq.~\eqref{eq:spec_delta} is the distance above the real axis. The spectral function obeys the sum rule $s=\int_{-\infty}^{\infty} \text{d} \omega \rho(\omega) = 1$, thus the Green's function has an asymptote $G(z) \rightarrow s/z$ for $|z| \to \infty$, according to Eq. \eqref{eq:hilbert}. 

Determining the spectral function $\rho(\omega)$ from a finite set of values $G_n$ with a finite precision is the goal of this paper. Sometimes one can divide the Green's function into a non-interacting Green's function and a correction part due to interactions, expressed by a self-energy function $\Sigma(z)$. Often one can determine the analytical continuation of the non-interacting Green's function exactly, and therefore it may be more convenient to perform the analytic continuation of the self-energy $\Sigma(z)$, instead of the full Green's function $G(z)$. The self-energy has the following analytical form:
\begin{equation}
\Sigma(z)=\Sigma_0+\Sigma_G(z)
\end{equation}
with a static part $\Sigma_0 \in \mathbb{R}$ and a dynamic part $\Sigma_G(z) \in \mathbb{C}$. The latter has the same analytic properties as $G(z)$ with the exception of the normalisation value, $s \in \mathbb{R}^+$, which is not necessarily equal to one~\cite{beach}. Once the self-energy has been analytically continued to the real axis, then the spectral function may be obtained via the Dyson equation and Eq.~\eqref{eq:spec_delta}. In general the procedure outlined above leads to smaller errors than performing directly the analytic continuation of $G(z)$~\cite{grechnev07prb76:035107,staar14,PhysRevB.80.045101}.

\section{\label{sec:pade}Pad\'{e} approximants}
A Pad\'{e} approximant can be expressed as a $[k/r]$ rational polynomial:
\begin{equation}
P_{k,r}(z)= \frac{\sum_{i=1}^{k+1} a_i z^{i-1} }{\sum_{i=1}^{r} b_i z^{i-1}+z^r},
\end{equation} 
and has in general complex coefficients $a_i$ and $b_i$, and an asymptote $a_{k+1}z^{k-r}$ for large $|z|$. Since $G(z)$ and $\Sigma_G(z)$ have asymptotes $s/z$, they are suitably fitted using a $[(r-1)/r]$ Pad\'{e} approximant, i.e. 
\begin{equation}
P_{r}(z)= \frac{\sum_{i=1}^{r} a_i z^{i-1} }{\sum_{i=1}^{r} b_i z^{i-1}+z^r}.
\end{equation} 
The coefficient $a_r=s \in \mathbb{R}^+$ is significant as it determines the asymptotic behaviour. The number of coefficients are in total $N=2r$ and can be found as described below. 

\subsection{Plain Pad\'{e}}
In Beach's algorithm~\cite{beach}, the $N$ coefficients for the Pad\'{e} approximant $P_r(z)$ of a complex function $f(z)$ are found by selecting $N$ points $z_i \in \mathbb{C}$ where $f(z)$ is known and requiring $P_r(z_i)=f(z_i)$. This results in inverting a $N$ by $N$ matrix. For sake of simplicity, let us call this approach Plain Pad\'{e}.

\subsection{\label{subsec:LSPade}Least Square (LS) Pad\'{e}}
Using the same number of unknown coefficients as fitting points may lead to overfitting and can give unstable continuations with spurious or even unphysical spectra, e.g. negative intensities. Hence, it may be more useful to have the number of input points $M$ and the number of coefficients $N$ independent under the condition $N\leqslant M$, as was also suggested by the authors in Ref.~\onlinecite{PhysRevB.87.245135}. 
Requiring $P_r(z_i)=f(z_i)$ for $M$ points, with $N\leqslant M$, yields a LS matrix equation
\begin{equation}
K\bm{v}=\bm{y},
\end{equation}  
where
\begin{equation}
\bm{v}=
\begin{bmatrix}
\mathbf{a} \\
\mathbf{b} 
\end{bmatrix}
= 
\begin{bmatrix}
a_1 \\
\vdots \\
a_{r} \\
b_1 \\
\vdots \\
b_{r} 
\end{bmatrix}, \quad
\bm{y}=
\begin{bmatrix}
z_1^r f(z_1) \\
z_2^r f(z_2) \\
\vdots \\
z_M^r f(z_M) 
\end{bmatrix} 
\end{equation}
and 
\begin{equation}
K=
\begin{bmatrix}
  	1            &\cdots & z_1^{r-1} & -f(z_1)   & \cdots & -f(z_1) z_1^{r-1}  \\
	1            &\cdots & z_2^{r-1} & -f(z_2)   & \cdots & -f(z_2) z_2^{r-1}  \\
  	\vdots    & \vdots & \vdots & \vdots        &  \vdots & \vdots \\
  	1           &\cdots & z_M^{r-1} & -f(z_M)  &  \cdots & -f(z_M) z_M^{r-1}  
\end{bmatrix}. 
 \end{equation}


\subsection{\label{subsec:average}Average Pad\'{e} approximants}
As mentioned in the introduction, for input data of high precision, the performance of Plain Pad\'{e} is excellent. When the precision is reduced, however, the quality of the continuation quickly degrades, and shows a strong dependence on the number $N$ of coefficients and input points. A small $N$ may lead to an approximant with not enough poles to reproduce the structure of the function $f(z)$. A big $N$ increases the risk of spurious poles coming from imperfect zero-pole pairs, leading to errors and unphysical features. Due to that the zero-pole pairs arise from numerical noise and have therefore a random distribution~\cite{Andreas}, we suggest to perform the analytic continuation for several $N$ and then average the spectra in contrast to just picking a single $N$-value. Moreover, for the LS Pad\'{e}, a similar argument can be formulated for the number of (independent) fitting points $M$. An average continuation is now defined by the interval of coefficients $N$ and the interval of input points $M$, as well as by the way the input points are picked. We use subsequent Matsubara points starting at $\omega_{n_0}$ (see Fig.\ref{figtikz} for explaination of $n_0$), even though other distributions are possible, like e.g. a logarithmic one. We can now identify a single continuation with the values of $\left \{ n_0, N, M  \right \}$, which we label with the configuration subscript $c$. The Pad\'{e} coefficients of $c$ are labeled as $\bm{v}_c$ and its spectrum as $\rho_c(\omega)=-1/\pi \text{Im}[P^{(c)}(\omega+i\delta)]$. We now have to decide how to vary the independent coefficients $\left \{ n_0, N, M  \right \}$. For simplicity we choose to use the following sets:
\begin{align}
n_0 & \in  \left \{    n_{0_\text{min}} {:} n_{0_\text{step}} {:} n_{0_\text{max}}  \right \}         \\
M & \in  \left \{    M_\text{min} {:} M_\text{step} {:} M_\text{max}     \right \}         \\
N  & \in  \left \{    N_\text{min} {:} N_\text{step} {:} \text{min}(N_\text{max},M)     \right \},    
\end{align}
where for example $n_{0_\text{min}} {:} n_{0_\text{step}} {:} n_{0_\text{max}}$ indicates $n_{0_\text{min}},n_{0_\text{min}}+n_{0_\text{step}},...,n_{0_\text{max}}$. 
Once we have computed the analytic continuations for all the parameters defined above, we can take a weighted average by summing over all configurations $c$, with weight $w_c$,
\begin{equation}
\rho(\omega)= \frac{1}{\sum_c w_c}\sum_c w_c \rho_c(\omega).
\end{equation} 
We have also considered averaging the Pad\'{e} approximant coefficients, instead of the spectra, for configurations with the same $N$ and then average the spectra from different $N$. However this turned out to give worse results and will not be discussed further in this article.

\begin{figure*}[]
\includegraphics[]{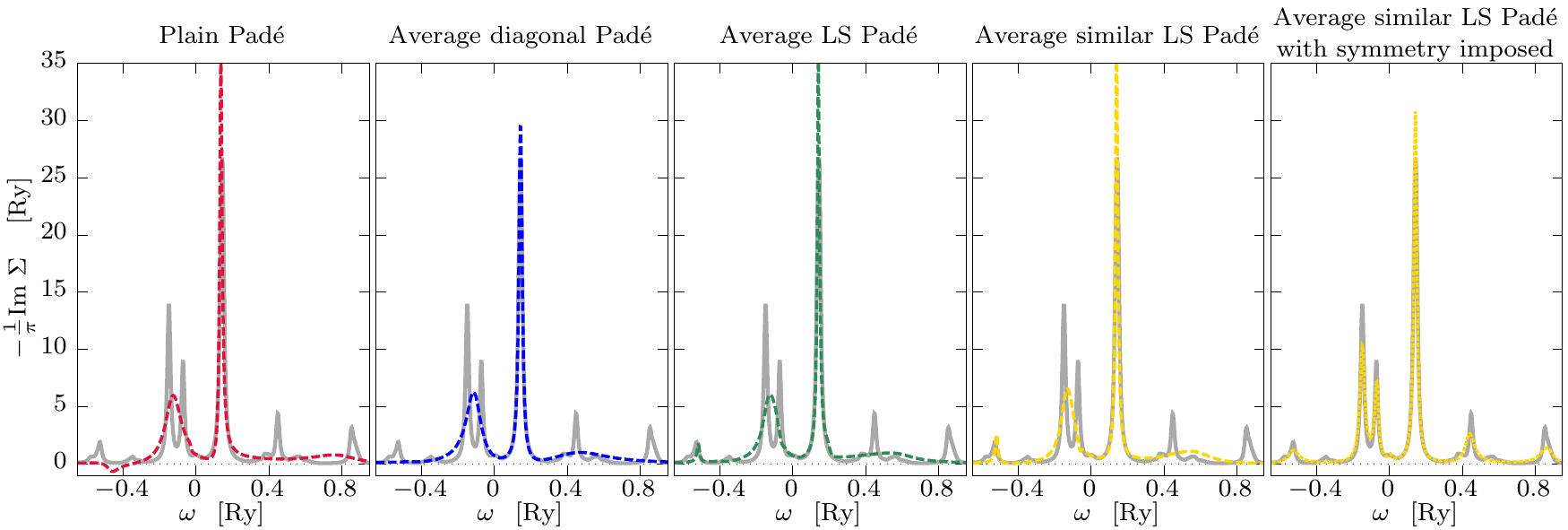}
\caption{(Color online) Illustration of the different improvements done on Plain Pad\'e for a Sm atom, with relative noise of magnitude $\sigma=10^{-6}$ on the Matsubara points. Description of the five obtained spectra (from left to right panel): Plain Pad\'e, using $N=M=70,n_0=1$, and characterised by unphysical features;  Pad\'{e} average with  $w_c^{D}$ and $n_0=1$; Pad\'{e} average with $w_c^{LS}$ and $n_0=1$;  Pad\'{e} average with $w_c^S$ and $n_0=1$; Pad\'{e} average with $w_c^S$ and with mirror symmetry imposed by using $n_0 \in \left \{-5{:}1{:}0 \right \}$.} 
\label{fig:Sm7:showImprovement:epslatex}
\end{figure*}

The next step is to specify the averaging weight distribution $w_c$. A necessary condition we apply to all considered distributions is the absence of unphysical configurations. A configuration $c$ is considered unphysical if $\rho_c(\omega)<0$ for some $\omega$. With this premise, we analyse three possible choices of distributions:
\subsubsection{Average diagonal Pad\'{e}}
 The distribution $w_c^{D}$ includes only physical configurations with $N=M$, i.e. $w_c^{D}=\delta_{N,M}$ for physical configurations and $w_c^{D}=0$ for unphysical configurations. The superscript $D$ refers to use of only diagonal configurations for which $N=M$.
\subsubsection{Average LS Pad\'{e}}
The distribution $w_c^{LS}$ includes all physical LS configurations (including those where $N=M$), hence $w_c^{LS}=0$ for unphysical configurations and $w_c^{LS}=1$ otherwise.
\subsubsection{Average similar LS Pad\'{e}}
The distribution $w_c^S$ includes a subset of all physical continuations. Spurious structures may arise due to the finite precision of the ill-conditioned matrix $K$ which needs to be inverted. Being randomly distributed, these spurious structures can be eliminated by favouring spectra similar to each other. To this aim we define a deviation-value between a physical spectral function from all the other physical ones:
\begin{equation}
\Delta_c= \sum_{c'\neq c} w_{c'}^{LS} \int_{-\infty}^{\infty} \text{d}\omega \left | \rho_c(\omega)-\rho_{c'}(\omega) \right |.
\end{equation} 
In order to filter out the best configurations, we introduce two criteria. Two sets of configurations will be generated, one for each fulfilled criterion, and we pick the intersection. One criterion is to only include configurations with $\Delta_c$ smaller than 30\% of the average deviation. The latter is given by
\begin{equation}
\bar{\Delta}=\frac{1}{\sum_c w_c^{LS}}\sum_c w_c^{LS} \Delta_c \: .   
\end{equation}
The other criterion is to restrict the number of configurations to be included to 51\% of all the physical ones. This is done by picking the ones with lowest $\Delta_c$. 

With the averaging weights, one can also calculate the energy resolved variation from the mean, with big variation for a particular energy indicating instability and possibly unresolved sharp features.

\subsection{\label{subsection:EnforcingSymmetry}Enforcing symmetry}
To improve the resulting continuation, it may be advantageous to exploit the symmetry $f(z^*)=f(z)^*$ by including negative Matsubara points, hence $n_0<1$, in the fitting. If a symmetric mesh is considered with equally many negative as positive Matsubara points ($n_0=-M/2+1$), spurious poles appear on the real axis, which results in a poor spectrum~\cite{PhysRevB.82.165125}. The more negative $n_0$ gets, the more spurious peaks appear. By averaging over several continuations many of the peaks will be suppressed due to their random positions. This makes it possible to have a more negative $n_0$-value compared to a non-averaging approach. By using a span of different $n_0$, the averaging procedure determines how many continuations are taken into account for each $n_0$. Assigning higher weights to continuations fulfilling the symmetry criterion $\text{Im}[a_r]=0$, in the spirit of Ref.~\onlinecite{beach}, sometimes improves on the accuracy even further, but this strategy has not been analysed in the present article.

\section{\label{sec:result}RESULTS AND DISCUSSION}
%
\subsection{\label{subsec:Improvement} Averaging methods}
In this study, we performed analytic continuation for a variety of functions. An interesting test for illustrating the averaging methods described in section~\ref{subsec:average} is offered by the self-energy of a Sm atom. This function is interesting for two reasons. First of all, it is a rather demanding test for a Pad\'{e} approximant method, due to the presence of several narrow peaks at close distance from each other, which are associated with the atomic multiplets (see grey lines in Fig.~\ref{fig:Sm7:showImprovement:epslatex}). Second, this Sm self-energy represents a realistic test case which may arise in state-of-the-art computational problems. The present function has been obtained by means of the electronic structure code RSPt~\cite{RSPtBook}, when addressing a cluster of seven Sm atoms in the Hubbard I approximation~\cite{PhysRevB.79.165104,grechnev07prb76:035107}. The Hubbard I approximation can be used to calculate the Green's function on the Matsubara axis. However, it can also be applied directly on the real axis which makes it possible to obtain a reference term to judge the quality of the analytic continuation. Further computational details about this test-case are described in Appendix \ref{appendix:ManyBody}. The spectral functions are evaluated at a distance $\delta=0.01$ Ry above the real axis. For the self-energy $\Sigma$, we estimate $\Sigma_0$ with $\tilde{\Sigma}_0$ just before doing the continuations, by fitting the asymptote of the real part to $\tilde{\Sigma}_0+c/\omega_n^2$. This gives us approximately $\Sigma_G$, which is suitably fitted by a $P[(r-1)/r]$ Pad\'{e} approximant described in section~\ref{sec:pade}. More sophisticated fittings of the asymptotic tail~\cite{granas12} have been tested, and lead to similar results. The dependence of the Pad\'{e} performance on the quality of the asymptotic fitting is analysed separately in Appendix~\ref{app:asymptotics}.

\begin{figure}[t!]
\includegraphics[]{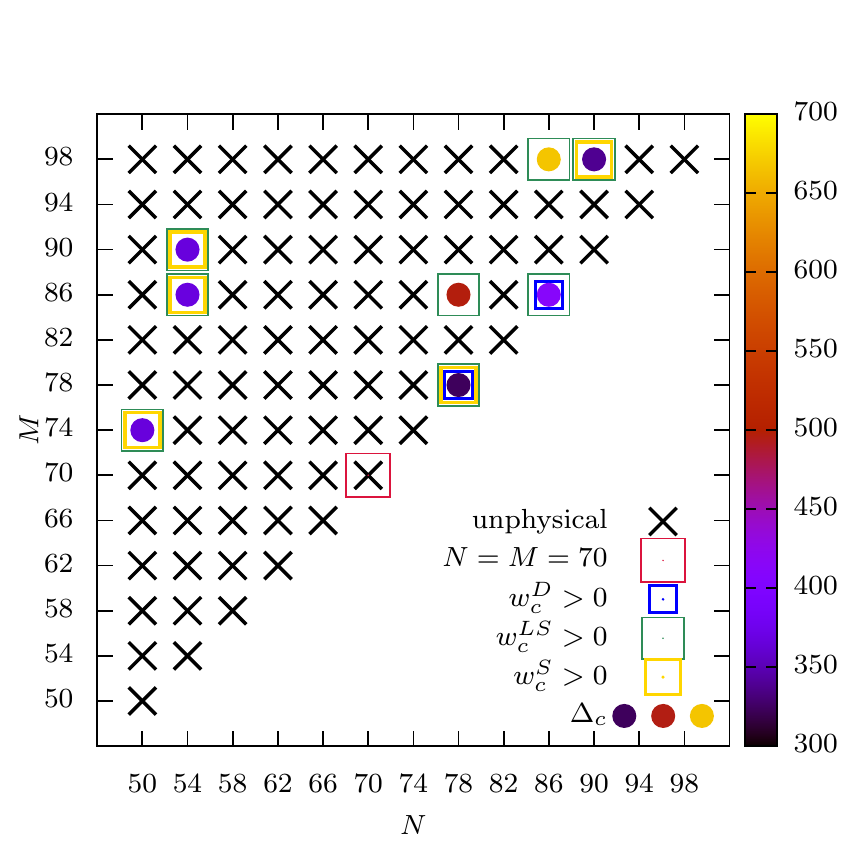}
\caption{(Color online) Illustration of configurations in $(N,M)$ space with $n_0=1$ for the self-energy of a Sm atom, with relative noise of magnitude $\sigma=10^{-6}$ on Matsubara points. A black cross denotes an unphysical, $\rho_c<0$, continuation.  A circular dot denotes a physical continuation, $\rho_c \geq 0$, and the colour denotes the deviation $\Delta_c$.}
\label{fig:NMspace}
\end{figure}

The Sm self-energy is continued by using the different averaging methods. The number of coefficients and input points to use in the average are set to $M \in  \left \{    50{:}4{:}98   \right \}$ and $N \in  \left \{    50{:}4{:}\text{min}(98,M)   \right \}$. The distributions of averaging weights and input points tested are instead the following: 
\begin{itemize}
\item Plain Pad\'{e}, $w_c=\delta_{M,70}\delta_{N,M}$ and $n_0=1$
\item Average diagonal Pad\'{e}, $w_c^{D}$ and $n_0=1$
\item Average LS Pad\'{e}, $w_c^{LS}$ and $n_0=1$
\item Average similar LS Pad\'{e}, $w_c^S$ and $n_0=1$
\item Average similar LS Pad\'{e} with mirror symmetry, $w_c^S$ and $n_0 \in  \left \{    -5{:}1{:}0   \right \} $
\end{itemize}
The results of the analytic continuations for these five different methods are shown in Fig.~\ref{fig:Sm7:showImprovement:epslatex}. The Matsubara self-energy has relative noise of magnitude $\sigma=10^{-6}$ on every Matsubara point. The dependence of the results on the noise level magnitude will be discussed in the next subsection, here we instead focus on the different averaging techniques. The leftmost panel of Fig.~\ref{fig:Sm7:showImprovement:epslatex} clearly illustrates that for this noise magnitude Plain Pad\'{e} leads to an unphysical self-energy, whose spectrum becomes negative at about -0.4 Ry. When averages are considered, this pathology is cured. Already with Average diagonal Pad\'{e}, i.e. by including only physical configurations with $N=M$, the self-energy is acceptable, as illustrated in the second panel (from the left) of Fig.~\ref{fig:Sm7:showImprovement:epslatex}. In the third panel, one can see the effect of averaging over more configurations with $N \leq M$. High energy peaks are now captured and the only significant deficiency is that the two peaks just below the Fermi level (zero energy) are merged into one. In the forth panel, among physical configurations those similar to each other are chosen. This does not change the spectrum for this test function case but it often gives an improvement versus the Average LS Pad\'{e} scheme if symmetry is imposed, which is discussed and shown in sec.~\ref{subsec:stability}. The distributions of configurations contributing to the averages for the first four cases are visualised in Fig.~\ref{fig:NMspace}. This figure shows a few physical continuations but it shall be noted that this is related to the particular test function as well as to the magnitude and representation of the external noise. For a generic function the number of physical continuations is typically larger. Moreover, the number of physical continuations can be increased by decreasing the steps $M_\text{step}$ and $N_\text{step}$. 
As a general consideration, it is not surprising that the Plain Pad\'{e} often runs into problems since so many continuations are not even physical. A final improvement to the spectral function is obtained when using the Average similar LS Pad\'{e} together with symmetry constraints. The rightmost figure of Fig.~\ref{fig:Sm7:showImprovement:epslatex} shows that this method gives an analytic continuation that is very close to the exact result. The two-peak structure at $-0.15$ Ry is finally resolved. The only noticeable differences are a certain tendency to broaden peaks which are close to each other. This broadening is not surprising and originates directly from the averaging. Not all the continuations contributing to the average are able to reproduce those peaks, and their inclusion results into a broadening.

\subsection{\label{subsec:stability}Stability with respect to the numerical noise}
The overview presented above is illustrative of the improvements due to the averaging in the Pad\'{e} scheme. It is important to analyse how these results depend on the numerical precision used in the computation. It is well-known that Beach's algorithm performs well for high precision problems~\cite{beach}. A fundamental question here is if only the precision of the input Matsubara data matters or if the internal numerics of Pad\'{e} also play a role, and to what extent. 

We show the influence of the Matsubara precision by illustrating how the analytic continuation of the Sm self-energy presented in the previous subsection changes by adding random noise of variable magnitude. For each Matsubara point, both the real and imaginary parts of the self-energy are multiplied with a factor $(1+\epsilon)$, where $\epsilon$ is a normally distributed random variable with zero mean and standard deviation value $\sigma$. Although the quality of a continuation is better judged by a close inspection of the resulting function with respect to the exact result, this procedure is difficult to follow for an assessement of a plethora of tests. To avoid this problem we follow Beach's strategy and devise a measure of the error performed on the spectral function:
\begin{equation}
E = \frac{\int_{-\infty}^\infty \text{d}\omega  \left |    \rho_\text{exact}(\omega)-\rho(\omega)    \right |    }{\int_{-\infty}^\infty \text{d}\omega  \left |    \rho_\text{exact}(\omega) \right | }.
\end{equation}
The integration bounds are of course limited in computations and were chosen as -1 Ry and +1 Ry, unless differently specified. To check how the error changes depending on noise representation, both mean $\left<E\right>$ and $\sigma_E=\left<(E-\left<E\right>)^2\right>$ are calculated using 10 different seeds for random number generation.

\begin{figure}[t]
\includegraphics[]{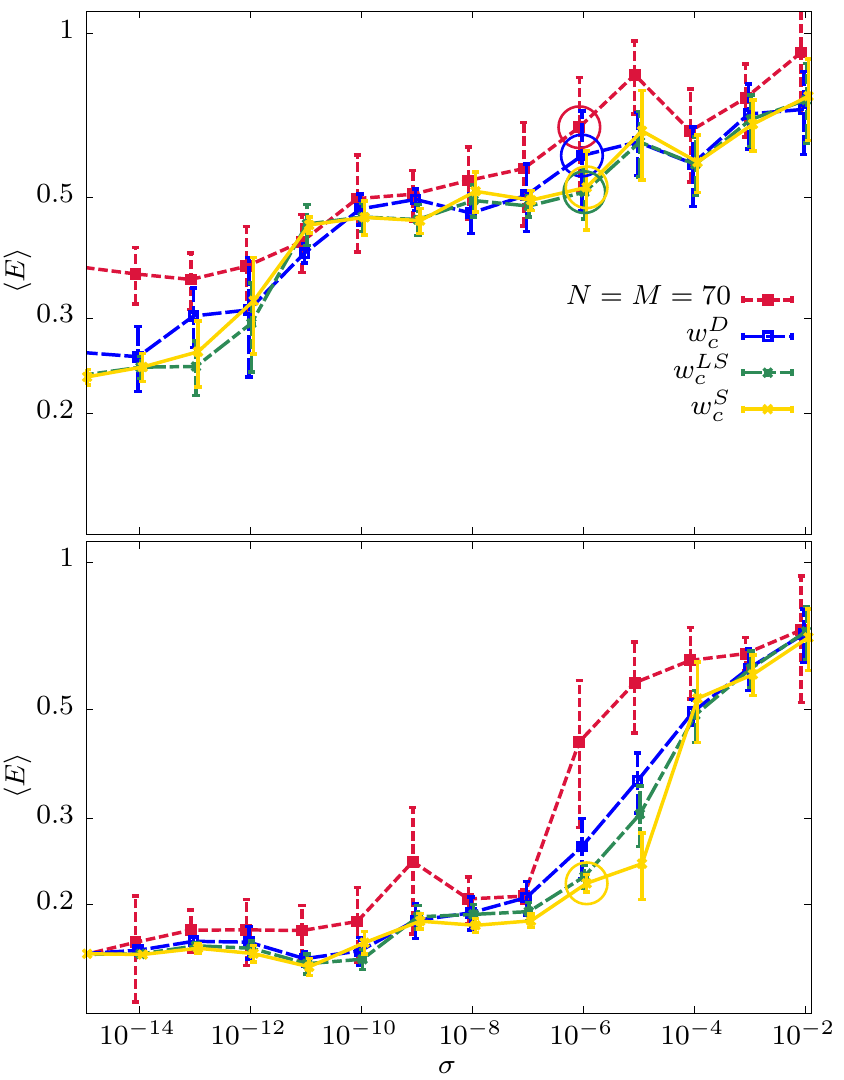}
\caption{(Color online) Average error $\left<E\right>$ versus standard deviation $\sigma$ of the numerical noise. The error deviation $\sigma_E$ (see main text) is represented as error bars. The five circles in the figure indicate those continuations corresponding to the spectra reported in Fig.~\ref{fig:Sm7:showImprovement:epslatex}. Top panel: analytic continuations performed with $n_0=1$. Bottom panel: analytic continuations performed with $n_0 \in  \left \{    -5{:}1{:}0   \right \}$, except for Plain Pad\'{e} in red where $n_0=-2$ is used.}
\label{fig:Sm:error:improve_new}
\end{figure}

In the top panel of Fig.~\ref{fig:Sm:error:improve_new} we report $\left<E\right>$ and $\sigma_E$ for the four different methods. Average diagonal Pad\'{e}, Average LS  Pad\'{e}, and Average similar LS  Pad\'{e} are considered by using the same setup as above but keeping $n_0=1$ for all of them. For Plain Pad\'{e} we use $N=M=70$, which is representative of a typical calculation. The circles in Fig.~\ref{fig:Sm:error:improve_new} indicate the errors $\left<E\right>$ associated to the spectra reported in Fig.~\ref{fig:Sm7:showImprovement:epslatex}. This may be useful to understand how the errors quantify different agreements with the exact spectral function. The data reported in the top panel of Fig.~\ref{fig:Sm:error:improve_new} illustrates how all the continuations using averaging leads to improved results over a wide range of noise magnitudes. Among the different averages, the schemes based on the LS minimisation perform better, but the improvements are minor. In the bottom panel of Fig.~\ref{fig:Sm:error:improve_new} we report analogous data, but using $n_0 \in  \left \{    -5{:}1{:}0   \right \}$ for all methods, except for the Plain Pad\'{e} where $n_0=-2$ is used. Three points become evident. First of all, introducing this small amount of negative frequencies leads to a great improvement of the continuations for noise of all magnitudes smaller than $\sigma=10^{-4}$. Second, the improvements due to this symmetrisation are much larger for the Pad\'{e} schemes based on averages, which again emphasises the limitations of the original Pad\'{e} formulation. Third, the curves for all different Pad\'{e} schemes based on averages exhibit some sort of plateaux when reducing the amount of noise. These are most likely due to the resolving power of the algorithms and will be analysed separately in section~\ref{sec:power}.

It is also instructing to look at the number of configurations included in $w_c^{D}, w_c^{LS}$ and $w_c^S$. These, resolved for each $n_0$, are reported in Table~\ref{table:n0} for a single noise representation of magnitude $\sigma=10^{-6}$. Notice that in our representation, $n_0=0$ corresponds to including the first negative Matsubara frequency, $n_0=-1$ corresponds to including also the second one, and so on. Interestingly, the three averaging methods give no solutions for $n_0=0$, although this is not a general feature of the method but specifically related to this function and noise. By using a lower $n_0$, the number of physical configurations increase. In principle, however, there is a non-trivial relation between physical configurations and $n_0$. For instance, we can see from Table~\ref{table:n0} that the averaging for $w_c^D$ picks up mostly configurations with $n_0=-3$ and $n_0=-5$.

Similar tests as those reported in Fig.~\ref{fig:Sm7:showImprovement:epslatex} have been performed for several other functions (not shown) and one can in general conclude that $w_c^{S}$ performs better than $w_c^{LS}$ which performs better than $w_c^D$. This hierarchy is also seen in Fig.~\ref{fig:Sm:error:improve_new}, although differences are smaller than compared to Plain Pad\'{e}. Finally notice, especially in the bottom figure, of Fig.~\ref{fig:Sm:error:improve_new} the big $\sigma_E$ for Plain Pad\'{e} compared to the other schemes. 

\bgroup
\def\arraystretch{1.2}
\begin{table}
\begin{tabular}{  l | l | l | l | l | l | l || l  } 
$n_0$                   &  -5   &  -4  &  -3 &  -2 &  -1  &  0  &  1    \\ \hline
$\# w_c^{D}>0$    & 13   &    8 & 11  &  3  &  1  &  0  &   2    \\ \hline
$\# w_c^{LS}>0$  &  91  &  75 & 81  & 44 &  3  &  0  &  8    \\ \hline
$\# w_c^S >0$      & 70  &  48 & 29  &   3 &   0 &  0  &   5    \\ 
\end{tabular}
\caption{Number of configurations contributing in averages, resolved for each $n_0$. Note that $w_c^S$ is determined by comparing separate sets of configurations with each other. In this case, all the configurations arising from $n_0 \in  \left \{    -5{:}1{:}0   \right \}$ are compared with each other. Configurations with $n_0=1$ are instead compared separately with each other. For each $n_0$, 91 configurations are in total considered and 13 configurations on the diagonal. The numbers in the rightmost column can be counted from Fig.~\ref{fig:NMspace}. The analysis is for a self-energy of a Sm atom, with $\sigma=10^{-6}$ noise added on the Matsubara points. }
\label{table:n0}
\end{table}
\egroup

\subsection{\label{subsec:inversion}Precision of the numerical routines}
\begin{figure}[t]
\includegraphics[]{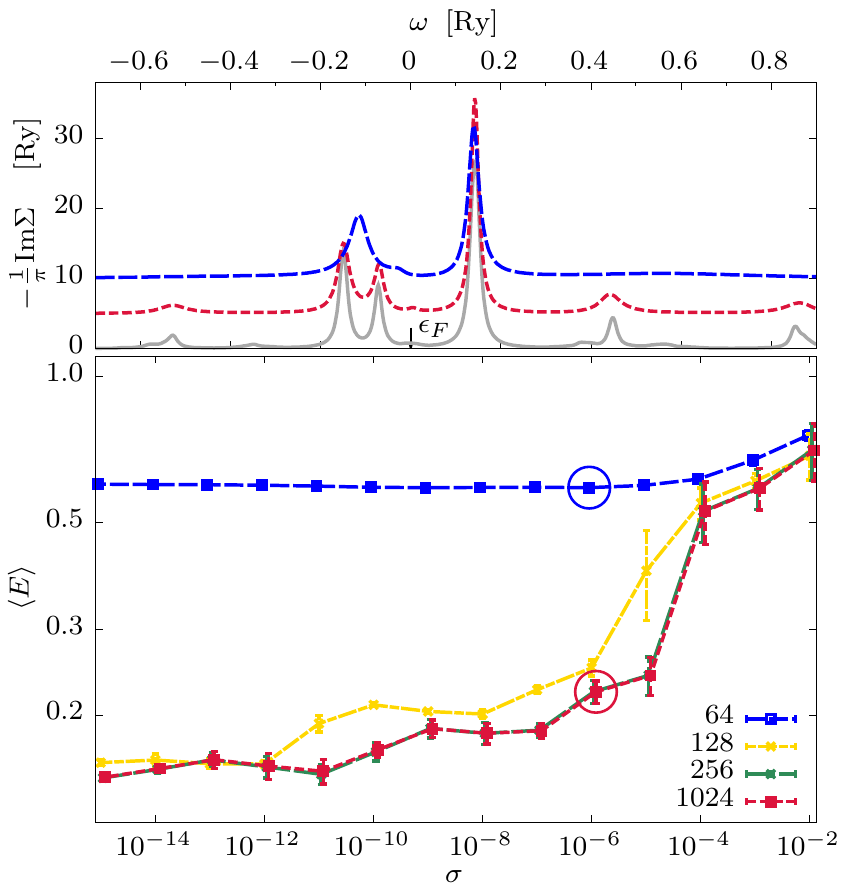} 
\caption{(Color online) In the bottom panel, the real axis error $\left<E\right>$ as a function of input noise $\sigma$ is reported for various precisions of the numerical routines (MPACK library). $\sigma_E$ is represented by the error bars. The circles indicate continuations whose spectra are plotted in the top panel, using the same lines as in the bottom panel, and shited with an offset for a better visualization. Exact results are also reported (lowest curve, in grey).}
\label{fig:inversion}      
\end{figure}

As mentioned above, an important question concerns the role of internal numerical precision of the Pad\'{e} approximant method. For Plain Pad\'{e}, i.e. Beach's original algorithm, one can estimate the needed precision for the inversion of the matrix $K$ by considering the ratio between the biggest and smallest elements, which is approximately given by $\xi=(w_{n_\text{max}})^r= \left (  (2(M+n_0-1)-1)\pi/\beta    \right )^r$.  This yields~\cite{PhysRevB.87.245135,beach} about $2 \log_2(\xi)$ binary digits, which in our setup varies between 73 and 256 depending on which $N,M$ and $n_0$ are used. From this one might expect 32 bits (single precision) and 64 bits (double precision) to perform poorly, and perhaps one should resort to 512 bits. To clarify these questions we perform several tests using the method which offered the best results for the Sm self-energy, i.e. Average similar LS Pad\'{e} with symmetry. We interface our code to a multiple precision arithmetic library MPACK~\cite{mpack}, using the standard routine \textit{CGESV}. We explore four different precisions of this routine: 64, 128, 256 and 1024 bits. The results of the tests are shown in Fig.~\ref{fig:inversion}. It is clear that standard double precision is not enough to solve the numerical problems arising in Pad\'{e}, unless a very big noise is present. Luckily, quadruple precision offers already good results. Increasing the precision even more, one obtains that 256 and 1024 bits give the same error, which means that no further improvements can be obtained. It is important to stress that these results depend on the choice of test function. In particular, a faster convergence to the exact continuation can be obtained for a more ``curvy'' function, without the delta-like peaks which are typical for atomic self-energies calculated with the Hubbard I approximation. Nevertheless,  Fig.~\ref{fig:inversion} illustrates how significant improvements to the analytic continuation can be obtained by using an increased precision for the LS-routines, even for big Matsubara noise.  We therefore use a modified LAPACK~\cite{laug} routine \textit{ZGELS} with a precision of 128 bits for the LS minimisation. In our tests LAPACK performed better than MPACK, and 128 bits LAPACK routines result into errors comparable to 256 bits MPACK. This most likely arises from using different algorithms for the LS problem. In light of these results the LAPACK routines with 128 bits precision were used for all the data presented in the rest of the paper, as well as for those presented in the previous sections.

\section{\label{sec:OtherAC}Comparison with other continuation methods}
In this section we compare the most accurate Pad\'{e} method, as described in the previous section, with other existing methods for analytical continuation. 
The latter include our in-house implementations of NNLS~\cite{NNLS}, NNT~\cite{Tikhonov} and MEM~\cite{PhysRevB.41.2380,Jarrell1996133,bryans,PhysRevB.44.6011,PhysRevE.81.056701,PhysRevE.82.026701,PhysRevB.81.155107,PhysRevB.82.165125}. A schematic overview of these approaches is given in appendix~\ref{app:others}. The test functions presented in the following were chosen to cover different types of analytic structures, from the multiplet structures of a Sm atom to the semi-circular density of states of the non-interacting infinite-dimensional Bethe lattice. Both Green's functions and self-energies are considered and the exact function is also known on the real axis to evaluate the accuracy and stability of the analytic continuation.
\begin{figure*}[t]
	\subfloat[Sm atom in a cluster]{\includegraphics[]{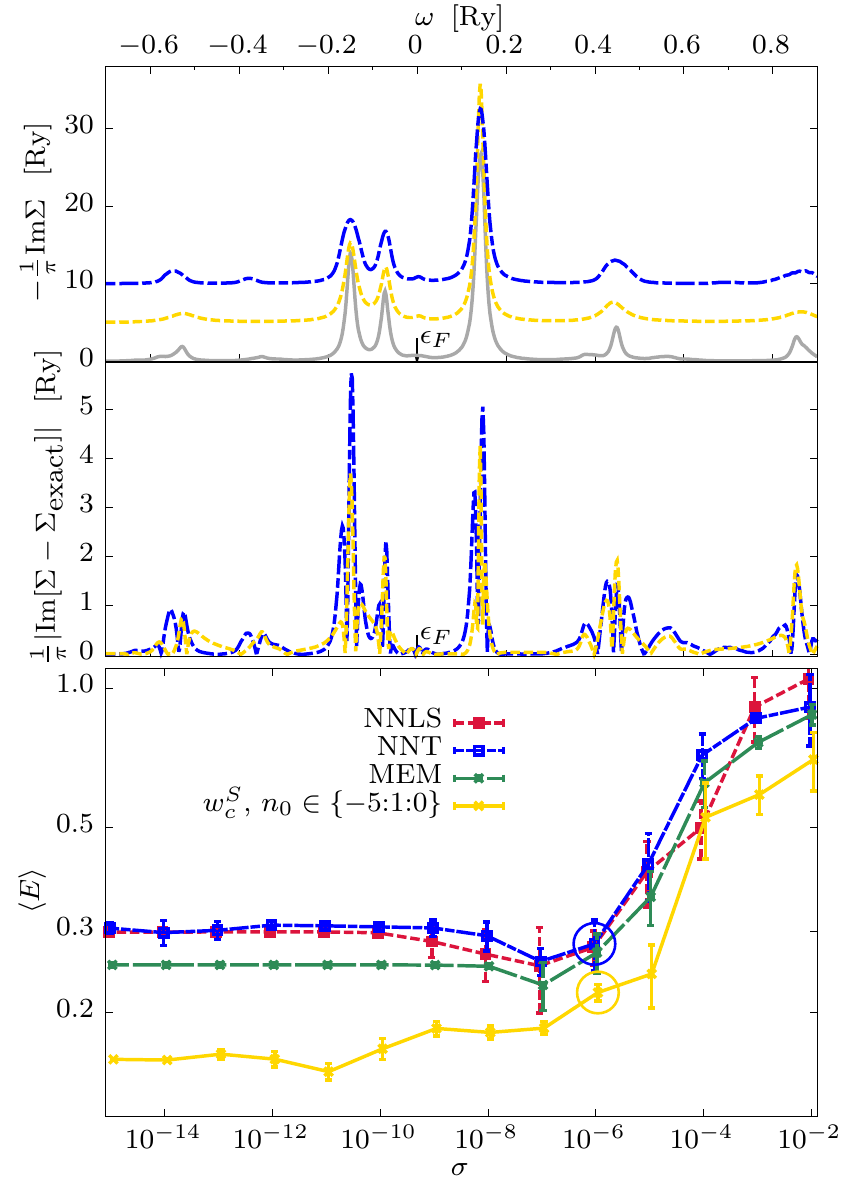}
		\label{fig:Smcluster}
	}
	\subfloat[Pr atom in the bulk]{\includegraphics[]{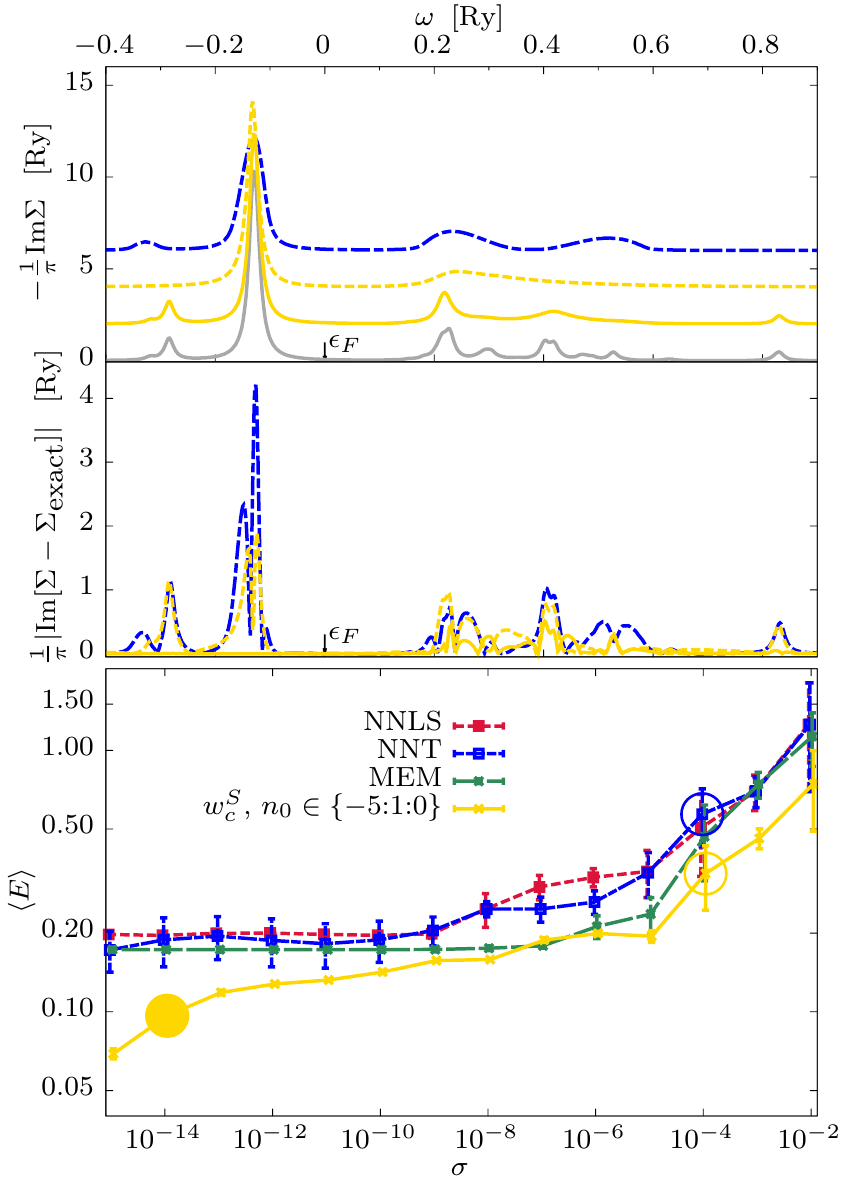}
		\label{fig:Prbulk}
	}
	\caption{(Color online) Comparison of the Pad\'e scheme with three other well-known methods for analytic continuation. In the bottom panels, the real axis error $\left<E\right>$ as a function of the Matsubara noise $\sigma$ is shown. The errors bars represent $\sigma_E$. The circles indicate continuations whose spectra are plotted in the top panels. Spectra shifted with an offset for a better visualisation. Exact results are also reported (lowest curves, in grey). In the middle panels the energy-resolved errors of the spectra in the top panels with respect to the exact result are reported.}
	\label{fig:Sm7andPrbulk}
\end{figure*}

\subsection{Sm atom in a cluster}
The first test case is represented by the function used in the previous sections, i.e. the atomic-like self-energy of one of the sites of a cluster of seven Sm atoms calculated in the Hubbard I approximation, as described in Appendix~\ref{appendix:ManyBody}. The error as a function of noise for several methods is presented on the bottom panel of Fig.~\ref{fig:Smcluster}. We note that the MEM gives a smaller error than NNLS and NNT, for most noise levels. Overall these three methods show a similar behaviour when varying the noise. The averaging Pad\'{e} scheme gives the most accurate results, except when $\sigma=10^{-4}$. Even in this case, nevertheless, the difference between Pad\'{e} and the best performing method is small. Finally, we note that none of the tested methods converge to the exact spectrum when decreasing noise up to the native input precision, and one can still observe a plateaux in $\left < E \right >$, as we emphasise in the section above. This confirms that this self-energy, with spiky features, is a difficult case for analytic continuation. The biggest contribution to the real-axis error for the two spectra in the top panel of Fig.~\ref{fig:Smcluster} originates from the underestimation of the peak heights for the three peaks closest to the Fermi-energy, which is clearly seen from the middle panel of Fig.~\ref{fig:Smcluster}.  

\subsection{Pr atom in the bulk}
The second test-case is given by the self-energy of a Pr atom, as obtained from LDA+DMFT simulations of bulk fcc Pr in the Hubbard I approximation (see also appendix~\ref{appendix:ManyBody}). The error as a function of noise for several methods is illustrated on the bottom panel of Fig.~\ref{fig:Prbulk}. As for Sm, we again see that the Pad\'{e} scheme outperforms the other methods for the majority of noise levels. For noise where $\sigma$ is between $10^{-5}$ and $10^{-8}$ the MEM shows a comparable agreement with the exact results. For big noise, it is very surprising that Pad\'{e} performs better than the MEM.
However, other tested self-energies based on the Hubbard I approximation, that are not shown in this article, also show the same behaviour.  
A significant difference between the left and right bottom panels of Fig.~\ref{fig:Sm7andPrbulk} is that for Pr no error plateaux is observed for the Pad\'{e} scheme, when decreasing noise. As explained in section~\ref{sec:power}, this is probably related to the analytic structure of the test function, i.e. the distribution of delta-like peaks within the considered energy range.  
From the upper and middle panel of Fig.~\ref{fig:Prbulk}, one can see that the advantage of the averaging Pad\'{e} scheme with respect to other methods is that, for comparable errors, the features it resolves are better captured, such as the height and width of the main peak.
This gives a better agreement with the exact continuation, even if more peaks are actually detected by other methods than by Pad\'{e}. 

\subsection{\label{subsec:bethe}Hubbard model on the Bethe lattice}
\begin{figure}[b]
\includegraphics[]{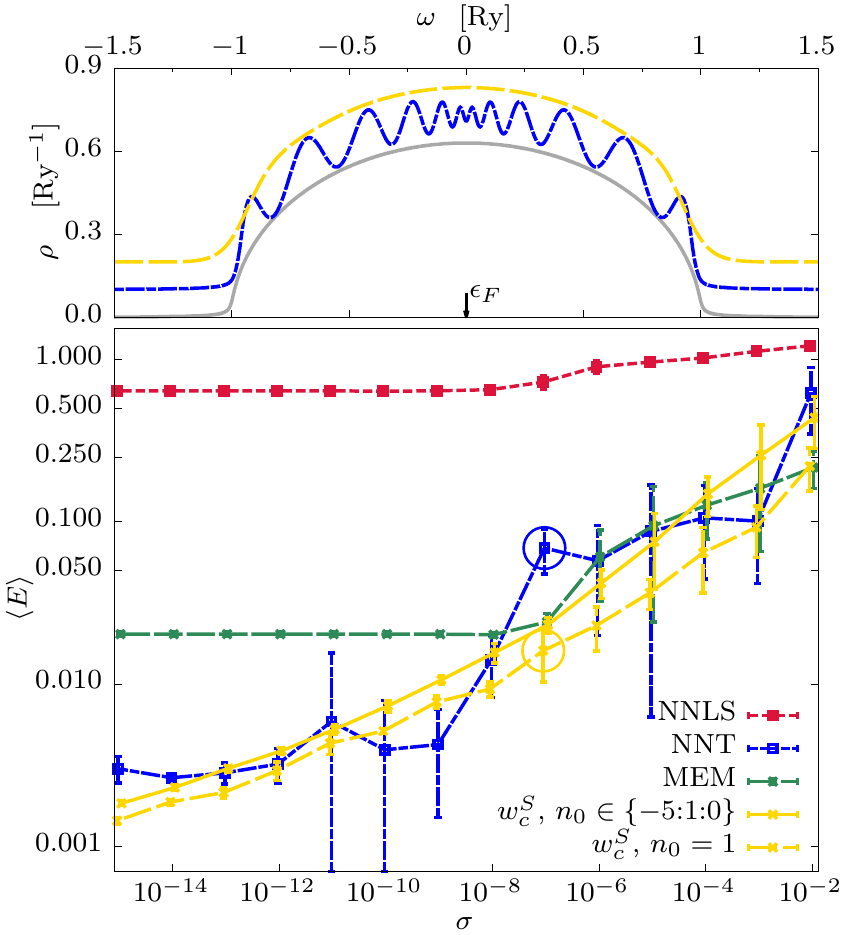}
\caption{(Color online) Comparing the Pad\'e scheme with three other well-known methods for the analytic continuation of the non-interacting Green's function of the Hubbard model on a Bethe-lattice with infinite nearest neighbors~\cite{georges96rmp68:13}. In the bottom panel, the real axis error $\left<E\right>$ as a function of the Matsubara noise $\sigma$ is shown. The error bars represent $\sigma_E$. The circles indicate continuations whose spectra are plotted in the top panel, using the same lines as in the bottom panel, and shifted with an offset for a better visualisation. Exact spectrum is also reported (lowest curve, in grey).}
\label{fig:Bethe:error:compare}
\end{figure}

An interesting test function is the non-interacting Green's function of the Hubbard model on the Bethe-lattice with infinite nearest neighbors~\cite{georges96rmp68:13}. This function has a smooth semi-circular shape, which makes it much different from the other functions considered so far. 
The error as a function of noise for $\beta=100$ Ry$^{-1}$ is shown in Fig.~\ref{fig:Bethe:error:compare} for several methods.
For the Pad\'{e} scheme, we see a stable monotonic decrease of $\left < E \right >$ as $\sigma$ is decreased. Even for big noise the error remains relatively small as a few spurious oscillations arise in the spectral function. The performance of the other methods varies greatly. The curve for the MEM shows no improvements for $\sigma$ smaller than $10^{-8}$, due to tiny oscillations around the exact function. We stress here that the MEM is known to have problems in describing sharp band edges far from the Fermi energy. The spectral function for NNT is also characterised by small spurious oscillations (see top panel of  Fig.~\ref{fig:Bethe:error:compare}), but they disappear with decreasing $\sigma$. The NNLS completely fails (at least in our implementation) in describing the present spectral function as many spurious peaks arise on the contour of the semi-circle.

 Another interesting observation from Fig.~\ref{fig:Bethe:error:compare} is that $n_0=1$ actually performs better than $n_0 \in \left \{ -5{:}1{:}0 \right \}$. The latter gives more spurious features, which makes it less precise. 
We have also obtained this result for some other test functions.
This trend we believe can be traced back to the presence of a discontinuity of the function when going from positive to negative Matsubara frequencies. If the function goes to zero as $i\omega$ goes to zero, like for the atomic self-energies of Sm and Pr, then adding a small number of negative frequencies improves the result significantly. 
However, if the function goes to a finite value, like for the metallic Green's function of the Bethe lattice, then trying to describe the discontinuity in the imaginary part at the real energy axis is detrimental to Pad\'{e}. This qualitative difference is not a problem for the Pad\'{e} method, since the asymptotic behaviour for small frequencies is known from the input Matsubara data, and one can decide beforehand whether to include negative frequencies or not. 
  

\subsection{Haldane model for a nano-ribbon}
\begin{figure}[t]
\includegraphics[]{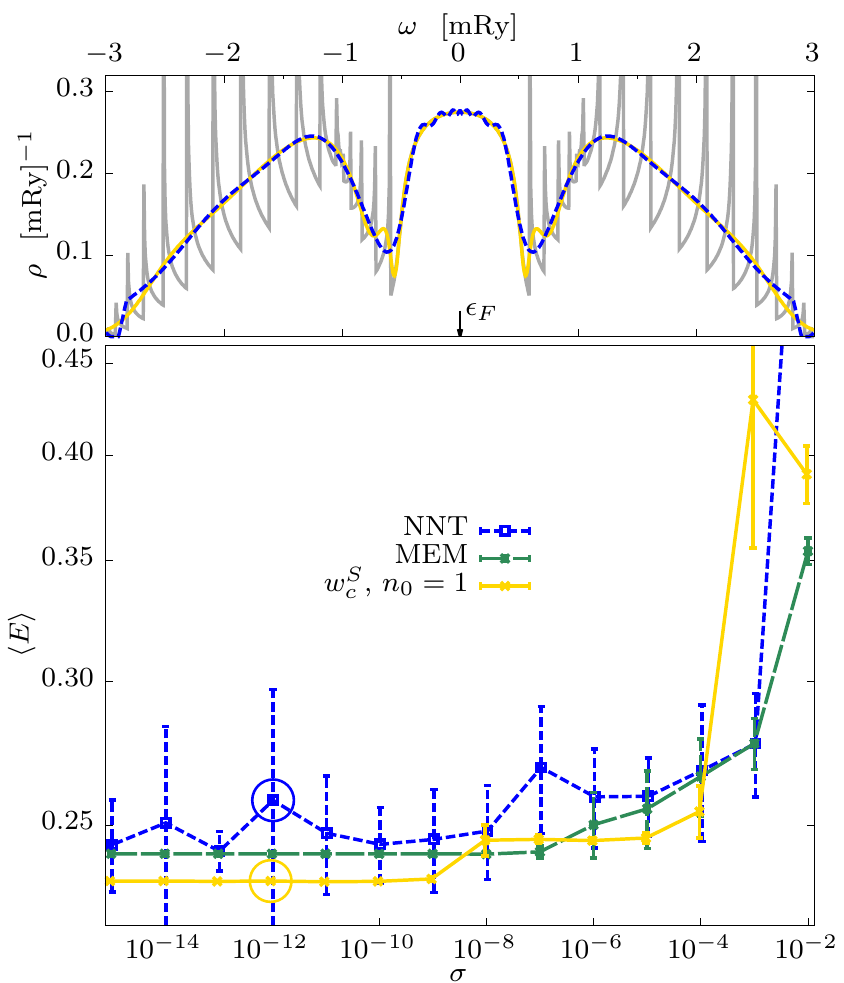}
\caption{(Color online) Comparing the Pad\'e scheme with three other well-known methods for the analytic continuation of the non-interacting Green's function of a nano-ribbon. In the bottom panel, the real axis error $\left<E\right>$ as a function of the Matsubara noise $\sigma$ is shown. The error bars represent $\sigma_E$. The circles indicate continuations whose spectra are plotted in the top panel, using the same lines as in the bottom panel. Exact spectrum is also reported (in grey).}
\label{fig:Haldane:NonInteracting:new_data:spectra:noise}
\end{figure}

We also investigate a non-interacting Green's function of an edge-atom of a nano-ribbon, having a honeycomb lattice for the Haldane model\cite{PhysRevLett.61.2015}. The spectral function contains many Van Hove singularities which makes it a very challenging test for analytic continuation. We use $\beta=100$ in units of mRy and a Hamiltonian hopping parameter ratio $\lambda/t=0.1$. In Fig.~\ref{fig:Haldane:NonInteracting:new_data:spectra:noise} real axis errors $\left < E \right >$ are shown for NNT, MEM and Pad\'{e}. We see that Pad\'{e}, NNT and MEM reproduce the exact low-energy spectrum successfully.  The overall distribution of spectral weight at higher energy is also captured. Resolving the high energy peaks require data of higher precision than those used here. NNLS creates a plethora of spurious peaks for all noise levels and its real axis error is therefore not shown.
This problem is related to the complicated structure of the exact function but also reflects the general tendency of NNLS to resolve a lot of features while leading to substantial errors on the peak positions. This behaviour is different from the present Pad\'{e} scheme. If the latter resolves the peak structure of a function the positions are usually correct. 

In the top panel of Fig.~\ref{fig:Haldane:NonInteracting:new_data:spectra:noise}, we see that the Pad\'{e} scheme reproduces the exact function very well at low energy and also the shoulders at $\pm 0.8$ mRy are captured. These are completely missed by the other methods. Overall, we again observe that the Pad\'e scheme leads to the best results among the tested methods. In light of the discussion at the end of section~\ref{subsec:bethe}, we use $n_0=1$ for this system, due to its metallic character.

\section{\label{sec:power}Algorithm resolving power}
\begin{figure}[]
\includegraphics[]{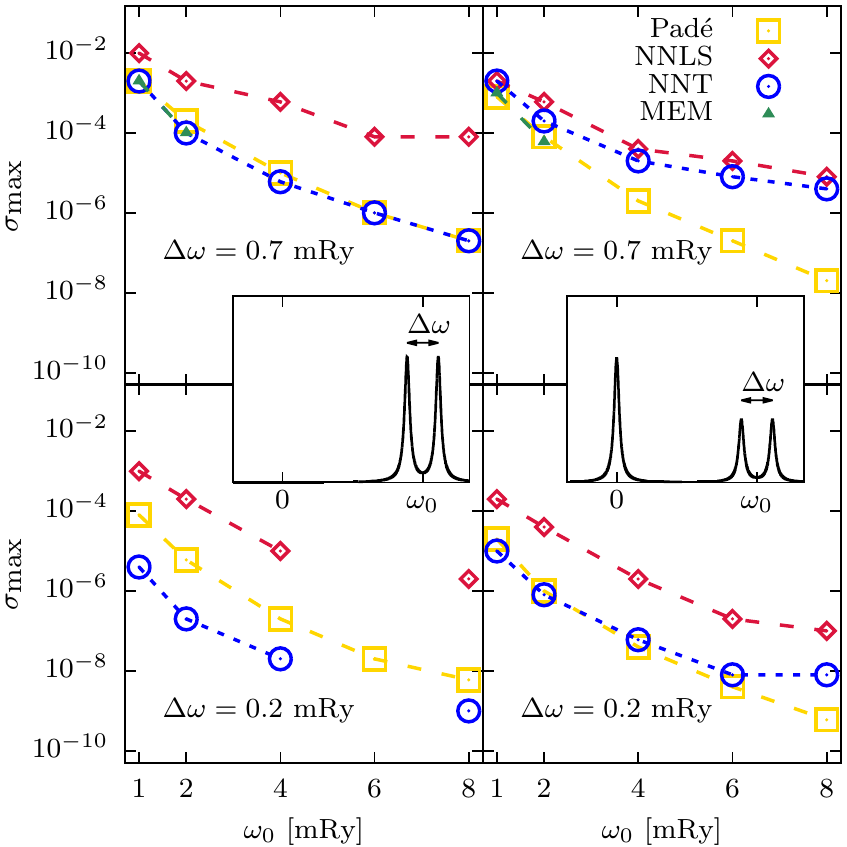}
\caption{(Color online) Left panel: highest possible noise magnitude to resolve the two-peak structure in Eq.~\ref{eq:twopeaksstructure} as a function of the center position $\omega_0$. Results for the Pad\'{e} scheme as well as NNLS, NNT and MEM are reported. Right panel: same as left panel but for a function with an additional peak at zero energy.}
\label{fig:Haverkort:omega0:4in1}
\end{figure}

From our previous analysis, it may be expected that decreasing the noise on the input Matsubara data should lead to continuous improvement of the analytical continuation obtained with Pad\'{e}. Nevertheless, the data reported on the bottom panel of Fig.~\ref{fig:Sm:error:improve_new} as well as in Fig.~\ref{fig:inversion} illustrate how below a certain noise no further decrease of $\left<E\right>$ is observed up to the native precision of the Matsubara data. This issue is related to the nature of the analytical continuation problem. The amount of information stored in the input data is not enough to resolve high energy features, even if one possesses the perfect continuation scheme. In this context, the Sm atom presented in the previous section is a particularly hard test, since there are several high energy satellites, like the double-peak at about -0.6 Ry seen in Fig.~\ref{fig:Sm7:showImprovement:epslatex}. The plateaux in $\left<E\right>$ can then be explained by the fact that the Pad\'{e} approximant method first manages to resolve all structures close to the Fermi level, but then saturates until the data precision is high enough to resolve also the high energy structures. This unfortunately does not happen in our test, see e.g. Fig.~\ref{fig:inversion}, as the native Matsubara data is limited to double precision. A similar problem occurs for the Green's function of the nano-ribbon, where none of the tested methods could resolve the series of high energy peaks. This discussion motivates us to analyse the ability of Pad\'{e} (and other methods) to resolve fine spectral structures, by using a simple test function. The latter is a Green's function which consists of two poles on the real axis, i.e.:
\begin{equation}\label{eq:twopeaksstructure}
G(z)=\frac{1/2}{z-(\omega_0+\Delta\omega/2)}+\frac{1/2}{z-(\omega_0-\Delta\omega/2)} \: .
\end{equation}
The spectrum is centered at $\omega_0$, while the two distinct peaks are located at a distance $\Delta\omega$ from each other. This Green's function is first generated at Matsubara frequencies for an inverse temperature $\beta=100$ mRy$^{-1}$. From the Matsubara data $G$ is continued to $\delta=0.1$ mRy above the real energy axis, using Pad\'e settings $w_c^S$ and $n_0 \in  \left \{    -5{:}1{:}0   \right \}$. Since we are treating a Green's function, no static part is subtracted from the asymptote. We can then identify the maximal noise magnitude $\sigma_\text{max}$ for which the two peaks can still be resolved. Success in resolving the two peaks is defined as the Pad\'{e} spectrum having two maxima, in a majority of 10 noise representations. In addition, the position of each maximum has to be within a distance $\delta$ from the position of the exact peak. The maximum allowed $\sigma_\text{max}$ for $\omega_0 \in \{ 1,2,4,6,8\}$ is reported in the left panels of Fig.~\ref{fig:Haverkort:omega0:4in1}, for two different distances $\Delta\omega$. If the two-peaks could not be resolved for any noise level, no point is plotted. As expected the accuracy of Pad\'{e} decreases when $\omega_0$ increases, and also when the peaks are closer to each other. It is also interesting to analyse how the algorithm performs when other spectral features are present. To this aim we add a pole centered at zero energy and containing half of the total spectral weight, see the right panels of Fig.~\ref{fig:Haverkort:omega0:4in1}. The resolving power of the algorithm is slightly decreased, but the scaling with respect to $\omega_0$ seems to be the same as for the original two-peak function. 

Before comparing to the other methods, let us discuss the scaling of the curves reported in  Fig.~\ref{fig:Haverkort:omega0:4in1}. The monotonic decrease of $\sigma_\text{max}$ as $\omega_0$ increases is expected and can be understood by comparing $G(z)$ with a single pole function: $1/(z-\omega_0)$ at $z=i\omega_n$. Expanding the difference of them in $\Delta \omega$ one gets $\Delta \omega^2/(i\omega_n-\omega_0)^3$, where it is clear the two-peak structure should be hard to resolve for a big $\omega_0$ or a small $\Delta \omega$. 
From Fig.~\ref{fig:Haverkort:omega0:4in1} we conclude it is less than exponentially hard for Pad\'e to resolve a two peak structure as a function of $\omega_0$. Overall, the data reported in Fig.~\ref{fig:Haverkort:omega0:4in1} offer a good quantitative measure of the resolving power of the Pad\'{e} scheme. When considering pairs of peaks at different distance from zero energy one can explain the real-axis error saturation as a function of Matsubara precision observed in the previous sections for the Pad\'{e} algorithm. 

The resolving power of the other methods shown in Fig.~\ref{fig:Haverkort:omega0:4in1} also suffers from the ill-posed nature of the analytic continuation. In addition, the saturation observed when decreasing Matsubara noise also comes from real-axis discretisation and regularisation constraints. In an overall comparison of the methods, the performance in resolving the two-peak structure is somehow the opposite to the performance for smooth curves, see e.g. subsection~\ref{subsec:bethe}. For the two-peak test function NNLS performs best, MEM worst, and NNT and the Pad\'e scheme are in the middle. This confirms what is outlined in the previous section, i.e. that NNLS resolves many more peaks than the other methods, although at the price of a significant loss of accuracy. Notice also how quickly the performance of MEM degrades with respect to the increase of $\omega_0$. Smearing of peaks by MEM is a well known issue\cite{PhysRevB.76.035115} and is here seen when comparing with the other analytic continuation methods. Finally, notice that only the Pad\'e scheme was able to resolve the peaks centered at $\omega_0=6$ mRy and separated by $\Delta\omega=0.2$ mRy. 

\section{\label{sec:conclusions}Conclusions}
We have developed a Pad\'{e} scheme to perform the analytic continuation of Green's functions and self-energies. We have first decoupled the number of input Matsubara points and the number of Pad\'{e} coefficients in Beach's algorithm. To improve the stability and the accuracy we propose to average several physical LS Pad\'{e} approximants as well as to use numerical routines of higher precision than the Matsubara data has. The averaging is shown to remove spurious features, arising due to the presence of imperfect zero-pole pairs and high numerical precision routines are shown to be crucial for resolving spectrum features, even for noisy Matsubara data. Enforcing mirror symmetry further improves the accuracy, for test functions with small spectral weight at zero energy. The joint usage of averaging physical LS continuations and mirror symmetry is shown to perform better than other well-known methods for analytic continuation, such as NNLS, NNT and MEM, for a variety of test functions. Even for noisy Matsubara data, the Pad\'{e} scheme is shown to give a better agreement with exact results than other methods. The problem of unphysical continuations, which is considered to be typical of Pad\'{e}, is here cured for all tested functions. The performance of our algorithm in resolving close peaks at high energies is also high, and inferior only to NNLS, among the investigated methods. But for functions with smooth and several peaky features, the Pad\'{e} scheme gives higher accuracy than NNLS.

An accurate Pad\'{e} scheme for analytic continuation presents several advantages with respect to other methods. 
The Pad\'{e} method assumes an ansatz for the shape of the exact function and does not depend on any further assumption, such as model functions, normalisation, regularisation or asymptote of the Matsubara data.  
This makes it tolerant to systematic errors. Moreover, spectral methods, such as MEM, need to discretise the real axis and sometimes the attained spectrum is very dependent on how this is done. This is not a problem for the Pad\'{e} method, where the function becomes known in the whole complex plane and can just be evaluated on the real energy window of interest. For the spectral methods, the energy window has to be chosen with greater care, since one wants a window which is small enough to have a dense mesh but big enough to ensure the whole spectral weight is included. A too small energy window does not only exclude spectral features but also reduces spectral accuracy inside, since left-out features have to be compensated. 

The developed Pad\'{e} scheme can find several applications in computational condensed matter physics. In LDA+DMFT it can be used to extract spectral functions when calculations are performed on the Matsubara axis. Moreover, increasing applications of the exact diagonalisation solver~\cite{thunstrom12prl109:186401,lu14prb90_085102,kolorenc12prb85_235136,granath12prb86:115111} demands for the development of new techniques to fit the hybridisation function more accurately.  One way to do this is to fit the function in the whole complex plane instead of using a selected axis (real or imaginary). This may be useful for codes working on the Matsubara axis, for which the exact diagonalisation fitting may be non-trivial~\cite{thunstrom12prl109:186401,kolorenc12prb85_235136}. Finally, several implementations of LDA+DMFT include analytic continuation in their computational schemes, such as e.g. the KKR/EMTO implementations of Refs.~\onlinecite{Andreas,PhysRevB.67.235106,minar05prb72:045125}. A more controlled formulation of Pad\'{e} like the one presented here may lead to an improvement of the stability and the accuracy of those codes.  

\section{\label{sec:acknowledgments}Acknowledgments}
This work was sponsored by the Swedish Research Council (VR), the Swedish strategic research programme eSSENCE and the Knut and Alice Wallenberg foundation (KAW). 
The computations were performed on resources provided by the Swedish National Infrastructure for Computing (SNIC).
We thank Andreas \"{O}stlin and Liviu Chioncel for insightful discussions, Maurits Haverkort for suggesting the section about resolving power, Andrea Droghetti for discussion and providing the input data for the Haldane test function and Erik van Loon for providing test functions and discussions.

\appendix

\section{\label{appendix:ManyBody}Atomic self-energies for Sm and Pr}
The self-energies of the Sm and Pr atoms used in the main text were calculated using DMFT within the Hubbard I approximation. In particular we used the RSPt code, where DMFT is interfaced with Full-Potential Linear Muffin-Tin Orbital (FP-LMTO) method~\cite{RSPtBook,PhysRevB.79.165104,granas12}. The aim of this appendix is to illustrate the basic concepts behind the calculations and also to present the computational details needed to reproduce our test functions. For an extensive explanation of methods and code, we redirect the reader to the aforementioned references.

In the Hubbard I approximation the many-body problem is reduced to an atomic problem, whose Hamiltonian is:
\begin{equation}
\hat{H}=\sum_{i,j}\hat{H}^{\text{at}}_{i,j}\hat{c}_{i}^\dagger  \hat{c}^{\phantom{\dagger}}_j  + \frac{1}{2}\sum_{i,j,k,l} U^{\phantom{\dagger}}_{i,j,k,l} \hat{c}_{i}^\dagger \hat{c}_j^\dagger  \hat{c}^{\phantom{\dagger}}_l \hat{c}^{\phantom{\dagger}}_k \:. 
\label{eq:Hamiltonian}
\end{equation}
Here the indices $i,j,k,l$ run over the correlated $4f$ orbitals, and the operators $\hat{c}$ and $\hat{c}^{\dagger}$ are respectively the annihilation and creation operators for those orbitals. The single particle atomic Hamiltonian matrix $\hat{H}^{\text{at}}_{i,j}$ and the Coulomb repulsion tensor $U_{i,j,k,l}$ fully determine the problem to solve. We assume the orbitals $i$ to be atomic-like, whose angular part corresponds to complex spherical harmonics $Y_{l,m}$ for $l=3$. In this representation, the Coulomb repulsion term is further parametrised using the Slater integrals $F^0$, $F^2$, $F^4$ and $F^6$, as explained in Ref.~\onlinecite{kotliar06rmp78:865}. For simplicity let us consider the chemical potential as being at zero energy. 

Once the Hamiltonian in Eq.~\ref{eq:Hamiltonian} is given, we can solve the atomic problem by direct diagonalisation, which leads to eigenvalues $E_{\nu}$ and eigenvectors $\ket{\nu}$. The Lehmann representation then gives the atomic interacting Green's function:
\begin{equation}
G_{i,j}(z) = \frac{1}{Z} \sum_{\nu_1,\nu_2} \frac{ \bra{\nu_1} c_i \ket{\nu_2}     \bra{\nu_2}c^\dagger_j \ket{\nu_1}     }{z+(E_{\nu_1}-E_{\nu_2})} (   e^{-\beta E_{\nu_1}} +e^{-\beta E_{\nu_2}}   ).
\end{equation}
From here, the self-energy can be obtained via inverse Dyson equation
\begin{equation}
\Sigma_{i,j}(z)=\left[\mathcal{G}_0^{-1}(z)\right]_{i,j}-\left[G^{-1}(z)\right]_{i,j} \: ,
\end{equation}
where the $\mathcal{{G}}_{0}$ is the non-interacting atomic Green's function.
In this article, we focus on the trace of the self-energy, i.e.:
\begin{equation}
\Sigma(z)= \text{Tr}[\Sigma_{i,j}(z)] = \sum_i \Sigma_{i,i}(z) \: .
\end{equation} 
The input points $z$ are chosen to be Matsubara frequencies for $\beta=100$~Ry$^{-1}$. The analytic continuation to the real energy axis is evaluated for a distance $\delta=0.01$~Ry.

The local Hamiltonians for a Sm atom in a cluster and a Pr atom in the bulk on the basis of the 14 correlated $4f$ orbitals are respectively  
\begin{widetext}
\setcounter{MaxMatrixCols}{14}
\begin{tiny}
\begin{equation*}
\begin{split}
{\text{\normalsize{Sm: \hspace{0.5cm} $\hat{H}^{\text{at}}_{i,j}$ =}}}&
\begin{pmatrix*}[r]
 -2.187 &  0 &  0 &  0 &  0 &  0 &  0 &  0 &  0 &  0 &  0 &  0 &  0 &  0 \\
  0 & -2.193 &  0 &  0 &  0 &  0 &  0 &  0.015 &  0 &  0 &  0 &  0 &  0 &  0 \\
  0 &  0 & -2.200 &  0 &  0 &  0 &  0 &  0 &  0.019 &  0 &  0 &  0 &  0 &  0 \\
  0 &  0 &  0 & -2.206 &  0 &  0 &  0 &  0 &  0 &  0.021 &  0 &  0 &  0 &  0 \\
  0 &  0 &  0 &  0 & -2.212 &  0 &  0 &  0 &  0 &  0 &  0.021 &  0 &  0 &  0 \\
  0 &  0 &  0 &  0 &  0 & -2.217 &  0 &  0 &  0 &  0 &  0 &  0.019 &  0 &  0 \\
  0 &  0 &  0 &  0 &  0 &  0 & -2.223 &  0 &  0 &  0 &  0 &  0 &  0.015 &  0 \\
  0 &  0.015 &  0 &  0 &  0 &  0 &  0 & -2.223 &  0 &  0 &  0 &  0 &  0 &  0 \\
  0 &  0 &  0.019 &  0 &  0 &  0 &  0 &  0 & -2.217 &  0 &  0 &  0 &  0 &  0 \\
  0 &  0 &  0 &  0.021 &  0 &  0 &  0 &  0 &  0 & -2.212 &  0 &  0 &  0 &  0 \\
  0 &  0 &  0 &  0 &  0.021 &  0 &  0 &  0 &  0 &  0 & -2.206 &  0 &  0 &  0 \\
  0 &  0 &  0 &  0 &  0 &  0.019 &  0 &  0 &  0 &  0 &  0 & -2.200 &  0 &  0 \\
  0 &  0 &  0 &  0 &  0 &  0 &  0.015 &  0 &  0 &  0 &  0 &  0 & -2.193 &  0 \\
  0 &  0 &  0 &  0 &  0 &  0 &  0 &  0 &  0 &  0 &  0 &  0 &  0 & -2.187 \\
\end{pmatrix*}  \\
\\
{\text{\normalsize{Pr: \hspace{0.5cm} $\hat{H}^{\text{at}}_{i,j}$ =}}}&
\begin{pmatrix*}[r] 
 -0.678 &  0 &  0 &  0 &  0 &  0 &  0 &  0 &  0 &  0 &  0 &  0 &  0 &  0 \\
  0 & -0.683 &  0 &  0 &  0 &  0 &  0 &  0.010 &  0 &  0 &  0 &  0 &  0 &  0 \\
  0 &  0 & -0.686 &  0 &  0 &  0 &  0 &  0 &  0.013 &  0 &  0 &  0 &  0 &  0 \\
  0 &  0 &  0 & -0.691 &  0 &  0 &  0 &  0 &  0 &  0.014 &  0 &  0 &  0 &  0 \\
  0 &  0 &  0 &  0 & -0.694 &  0 &  0 &  0 &  0 &  0 &  0.014 &  0 &  0 &  0 \\
  0 &  0 &  0 &  0 &  0 & -0.699 &  0 &  0 &  0 &  0 &  0 &  0.013 &  0 &  0 \\
  0 &  0 &  0 &  0 &  0 &  0 & -0.702 &  0 &  0 &  0 &  0 &  0 &  0.010 &  0 \\
  0 &  0.010 &  0 &  0 &  0 &  0 &  0 & -0.702 &  0 &  0 &  0 &  0 &  0 &  0 \\
  0 &  0 &  0.013 &  0 &  0 &  0 &  0 &  0 & -0.699 &  0 &  0 &  0 &  0 &  0 \\
  0 &  0 &  0 &  0.014 &  0 &  0 &  0 &  0 &  0 & -0.694 &  0 &  0 &  0 &  0 \\
  0 &  0 &  0 &  0 &  0.014 &  0 &  0 &  0 &  0 &  0 & -0.691 &  0 &  0 &  0 \\
  0 &  0 &  0 &  0 &  0 &  0.013 &  0 &  0 &  0 &  0 &  0 & -0.686 &  0 &  0 \\
  0 &  0 &  0 &  0 &  0 &  0 &  0.010 &  0 &  0 &  0 &  0 &  0 & -0.683 &  0 \\
  0 &  0 &  0 &  0 &  0 &  0 &  0 &  0 &  0 &  0 &  0 &  0 &  0 & -0.678 \\
\end{pmatrix*} 
\end{split}
\end{equation*}
\end{tiny}
\end{widetext}
The orbitals are ordered from $m=-3$ to $m=3$ for minority spin and then again for majority spin, for both rows and columns. 
The Slater parameters for determining the Coulomb tensor are instead
\begin{eqnarray*}
\text{F}^0= 0.441~\text{Ry} & \qquad \text{F}^2= 0.876~\text{Ry} \\
\text{F}^4= 0.585~\text{Ry} & \qquad \text{F}^6= 0.433~\text{Ry}
\end{eqnarray*}
for Sm and
\begin{eqnarray*}
\text{F}^0= 0.514~\text{Ry} & \qquad \text{F}^2= 0.745~\text{Ry} \\
\text{F}^4= 0.488~\text{Ry} & \qquad \text{F}^6= 0.360~\text{Ry}
\end{eqnarray*}
for Pr. As a final remark, please consider that these two tests should be intended as realistic examples but not as physical solutions of given problems.

\section{Dependence on the static self-energy asymptote} \label{app:asymptotics}
\begin{figure}[]
        \includegraphics[]{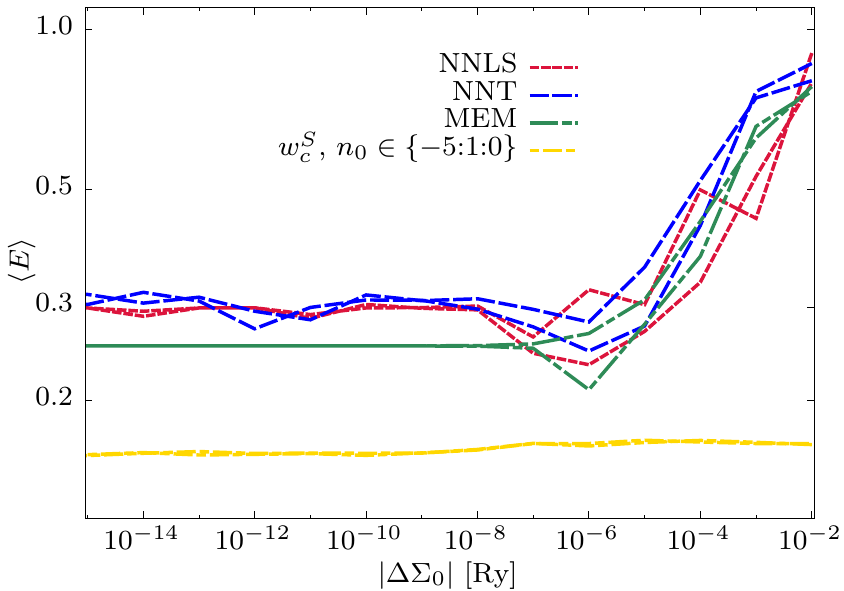}
        \caption{(Color online) Real-axis error as a function of the residual static self-energy $\Delta \Sigma_0$. Both positive and negative $\Delta \Sigma_0$-values are considered, hence the two lines per method.  The test is for the self-energy of a Sm atom with double precision Matsubara data.}
        \label{staticSigma}
\end{figure}

In this appendix we will investigate how sensitive the different continuation methods are to a small static part in the self-energy, which might arise from imprecise asymptotic fitting. 
The self-energy can be decomposed as $\Sigma=\Sigma_0+\Sigma_G$, where $\Sigma_G(z) \to 0$ as $|z| \to \infty$. All the considered continuation methods work with $\Sigma_G$ and assume that it has the correct asymptote.  
However, since we estimate $\Sigma_0$ by fitting a constant $\tilde{\Sigma}_0$ to the noisy data, the function to continue may have a small static part $\Delta \Sigma_0 = \Sigma_0-\tilde{\Sigma}_0$. Therefore, it is interesting to test how sensitive the continuation methods are to a residual static part.
We perform this test for the self-energy of a Sm atom, which is described extensively in the main text. We first subtract the static part $\Sigma_0$ from the raw data, then add a small static part $\Delta \Sigma_0$ and perform the analytic continuation. Finally we calculate the real-axis error, by comparing to the exact self-energy, which is reported in Fig.~\ref{staticSigma}. It is clear that the Pad\'e algorithm is much less sensitive to a finite $\Delta \Sigma_0$ than the other methods. This is typical of the method and not only of our particular Pad\'{e} scheme. This tolerance to errors on the asymptote is one of the reasons why the Pad\'e scheme works so well also for big noise magnitudes compared with the other methods.     
Note that Green's functions do not have a finite static part, but this test may be relevant for them too, since systematic errors in numerical routines (e.g. Fourier transforms) may affect their asymptotic behaviour.

\section{Overview of other methods for analytic continuation}\label{app:others}
In this appendix we provide a brief description of the other methods for analytic continuation that are used in section~\ref{sec:OtherAC}. For a more detailed description of all techniques mentioned here, we redirect the reader to appropriate references. Note that in this implementation we use standard double precision (64 bits). In a few cases some continuations improved by increasing the precision to quadruple (128 bits) but we have not investigated this issue in detail, since it is too far from the scope of this work.

\subsection{\label{subsec:NNLS}Non-Negative Least Square (NNLS)}
Instead of fitting to input Matsubara data $f_n$ using a Pad\'{e} approximant, one can use the Hilbert transform, Eq. \eqref{eq:hilbert}, to calculate the unknown $\rho(\omega)$. By discretising the real axis, the integral in Eq. \eqref{eq:hilbert} becomes
\begin{equation}
f_n=\sum_{j=1}^{N}  \frac{w_j}{i\omega_n-\omega_j} \rho_j \: ,
\end{equation}  
where $w_j$ is an integration weight and $\rho_j=\rho(\omega_j)$. The previous equation can be conveniently written as matrix equation of the form $\mathbf{f}=K \bm{\rho}$.
By choosing $M$ Matsubara points, such that $M \geq 2N$, and by respecting the non-negativity of the spectral function, the task becomes solving the following NNLS problem
\begin{equation}
\label{eq:NNLS}
\underset{\bm{\rho}\geq 0}{\text{min}} \left | \bm{f}-K\bm{\rho} \right |^2.
\end{equation}
A procedure to solve NNLS problems iteratively is described extensively in Ref.~\onlinecite{NNLS}.

\subsection{\label{subsec:NNT}Non-Negative Tikhonov (NNT)}
Ill-posed problems, such as analytic continuation, are often tackled by introducing regularisations. The Tikhonov regularisation~\cite{L-curve} is based on adding a minimisation condition on $\rho$, which results into transforming Eq.~\eqref{eq:NNLS} into
\begin{equation}
\label{eq:NNT}
\underset{\bm{\rho}\geq 0}{\text{min}} \left | \bm{f}-K\bm{\rho} \right |^2+\left |\alpha\bm{\rho} \right |^2=\underset{\bm{\rho}\geq 0}{\text{min}} \left | \tilde{\bm{f}}-\tilde{K}_\alpha\bm{\rho} \right |^2 \: .
\end{equation}
For a fixed $\alpha$ this is simply a NNLS problem and a suitable $\alpha$ can be determined by e.g. the L-curve method~\cite{L-curve}.

\subsection{\label{subsec:MEM}Maximum Entropy Method (MEM)}
The MEM is the most widespread method for analytic continuation~\cite{PhysRevB.41.2380,Jarrell1996133,bryans,PhysRevB.44.6011,PhysRevE.81.056701,PhysRevE.82.026701,PhysRevB.81.155107,PhysRevB.82.165125}. Its success is due to the robustness to Matsubara noise. The lack of information on the noisy Matsubara data is counteracted by additional information provided through a model function $\bm{m}$. In the MEM the spectral function is regularised by maximising the entropy. It is a Bayesian approach with a normal probability distribution as a likelihood function and a priori distribution set by the Shannon entropy. In the simplified case where the Matsubara data is assumed to be uncorrelated and with equal precision, the equation to solve can be formulated as:
\begin{equation}
\underset{\bm{\rho}\geq 0}{\text{min}} \left | \bm{f}-K\bm{\rho} \right |^2- \alpha S[\bm{\rho}] , 
\end{equation}
where $S$ is the entropy which is maximised for $\bm{\rho=m}$.

\bibliography{bibtex}{}

\begin{thebibliography}{46}%
\makeatletter
\providecommand \@ifxundefined [1]{%
 \@ifx{#1\undefined}
}%
\providecommand \@ifnum [1]{%
 \ifnum #1\expandafter \@firstoftwo
 \else \expandafter \@secondoftwo
 \fi
}%
\providecommand \@ifx [1]{%
 \ifx #1\expandafter \@firstoftwo
 \else \expandafter \@secondoftwo
 \fi
}%
\providecommand \natexlab [1]{#1}%
\providecommand \enquote  [1]{``#1''}%
\providecommand \bibnamefont  [1]{#1}%
\providecommand \bibfnamefont [1]{#1}%
\providecommand \citenamefont [1]{#1}%
\providecommand \href@noop [0]{\@secondoftwo}%
\providecommand \href [0]{\begingroup \@sanitize@url \@href}%
\providecommand \@href[1]{\@@startlink{#1}\@@href}%
\providecommand \@@href[1]{\endgroup#1\@@endlink}%
\providecommand \@sanitize@url [0]{\catcode `\\12\catcode `\$12\catcode
  `\&12\catcode `\#12\catcode `\^12\catcode `\_12\catcode `\%12\relax}%
\providecommand \@@startlink[1]{}%
\providecommand \@@endlink[0]{}%
\providecommand \url  [0]{\begingroup\@sanitize@url \@url }%
\providecommand \@url [1]{\endgroup\@href {#1}{\urlprefix }}%
\providecommand \urlprefix  [0]{URL }%
\providecommand \Eprint [0]{\href }%
\providecommand \doibase [0]{http://dx.doi.org/}%
\providecommand \selectlanguage [0]{\@gobble}%
\providecommand \bibinfo  [0]{\@secondoftwo}%
\providecommand \bibfield  [0]{\@secondoftwo}%
\providecommand \translation [1]{[#1]}%
\providecommand \BibitemOpen [0]{}%
\providecommand \bibitemStop [0]{}%
\providecommand \bibitemNoStop [0]{.\EOS\space}%
\providecommand \EOS [0]{\spacefactor3000\relax}%
\providecommand \BibitemShut  [1]{\csname bibitem#1\endcsname}%
\let\auto@bib@innerbib\@empty
\bibitem [{\citenamefont {Bednorz}\ and\ \citenamefont
  {M{\"u}ller}(1986)}]{highTc_ref}%
  \BibitemOpen
  \bibfield  {author} {\bibinfo {author} {\bibfnamefont {J.}~\bibnamefont
  {Bednorz}}\ and\ \bibinfo {author} {\bibfnamefont {K.}~\bibnamefont
  {M{\"u}ller}},\ }\href {\doibase 10.1007/BF01303701} {\bibfield  {journal}
  {\bibinfo  {journal} {Zeitschrift f{\"u}r Physik B Condensed Matter}\
  }\textbf {\bibinfo {volume} {64}},\ \bibinfo {pages} {189} (\bibinfo {year}
  {1986})}\BibitemShut {NoStop}%
\bibitem [{\citenamefont {Jonker}\ and\ \citenamefont
  {Santen}(1950)}]{Jonker1950337}%
  \BibitemOpen
  \bibfield  {author} {\bibinfo {author} {\bibfnamefont {G.}~\bibnamefont
  {Jonker}}\ and\ \bibinfo {author} {\bibfnamefont {J.~V.}\ \bibnamefont
  {Santen}},\ }\href {\doibase http://dx.doi.org/10.1016/0031-8914(50)90033-4}
  {\bibfield  {journal} {\bibinfo  {journal} {Physica}\ }\textbf {\bibinfo
  {volume} {16}},\ \bibinfo {pages} {337 } (\bibinfo {year}
  {1950})}\BibitemShut {NoStop}%
\bibitem [{\citenamefont {Georges}\ \emph {et~al.}(1996)\citenamefont
  {Georges}, \citenamefont {Kotliar}, \citenamefont {Krauth},\ and\
  \citenamefont {Rozenberg}}]{georges96rmp68:13}%
  \BibitemOpen
  \bibfield  {author} {\bibinfo {author} {\bibfnamefont {A.}~\bibnamefont
  {Georges}}, \bibinfo {author} {\bibfnamefont {G.}~\bibnamefont {Kotliar}},
  \bibinfo {author} {\bibfnamefont {W.}~\bibnamefont {Krauth}}, \ and\ \bibinfo
  {author} {\bibfnamefont {M.~J.}\ \bibnamefont {Rozenberg}},\ }\href@noop {}
  {\bibfield  {journal} {\bibinfo  {journal} {Rev. Mod. Phys.}\ }\textbf
  {\bibinfo {volume} {68}},\ \bibinfo {pages} {13} (\bibinfo {year}
  {1996})}\BibitemShut {NoStop}%
\bibitem [{\citenamefont {Kotliar}\ \emph {et~al.}(2006)\citenamefont
  {Kotliar}, \citenamefont {Savrasov}, \citenamefont {Haule}, \citenamefont
  {Oudovenko}, \citenamefont {Parcollet},\ and\ \citenamefont
  {Marianetti}}]{kotliar06rmp78:865}%
  \BibitemOpen
  \bibfield  {author} {\bibinfo {author} {\bibfnamefont {G.}~\bibnamefont
  {Kotliar}}, \bibinfo {author} {\bibfnamefont {S.~Y.}\ \bibnamefont
  {Savrasov}}, \bibinfo {author} {\bibfnamefont {K.}~\bibnamefont {Haule}},
  \bibinfo {author} {\bibfnamefont {V.~S.}\ \bibnamefont {Oudovenko}}, \bibinfo
  {author} {\bibfnamefont {O.}~\bibnamefont {Parcollet}}, \ and\ \bibinfo
  {author} {\bibfnamefont {C.~A.}\ \bibnamefont {Marianetti}},\ }\href
  {\doibase 10.1103/RevModPhys.78.865} {\bibfield  {journal} {\bibinfo
  {journal} {Rev. Mod. Phys.}\ }\textbf {\bibinfo {volume} {78}},\ \bibinfo
  {pages} {865} (\bibinfo {year} {2006})}\BibitemShut {NoStop}%
\bibitem [{\citenamefont {Aryasetiawan}\ and\ \citenamefont
  {Gunnarsson}(1998)}]{aryasetiawan98rmp61:237}%
  \BibitemOpen
  \bibfield  {author} {\bibinfo {author} {\bibfnamefont {F.}~\bibnamefont
  {Aryasetiawan}}\ and\ \bibinfo {author} {\bibfnamefont {O.}~\bibnamefont
  {Gunnarsson}},\ }\href {http://stacks.iop.org/0034-4885/61/i=3/a=002}
  {\bibfield  {journal} {\bibinfo  {journal} {Reports on Progress in Physics}\
  }\textbf {\bibinfo {volume} {61}},\ \bibinfo {pages} {237} (\bibinfo {year}
  {1998})}\BibitemShut {NoStop}%
\bibitem [{\citenamefont {Titchmarsh}(1939)}]{unique}%
  \BibitemOpen
  \bibfield  {author} {\bibinfo {author} {\bibfnamefont {C.}~\bibnamefont
  {Titchmarsh}},\ }\href@noop {} {\emph {\bibinfo {title} {The Theory of
  Functions}}},\ Fundamental Theories of Physics\ (\bibinfo  {publisher}
  {Oxford University Press, London, England},\ \bibinfo {year}
  {1939})\BibitemShut {NoStop}%
\bibitem [{\citenamefont {Silver}\ \emph {et~al.}(1990)\citenamefont {Silver},
  \citenamefont {Sivia},\ and\ \citenamefont {Gubernatis}}]{PhysRevB.41.2380}%
  \BibitemOpen
  \bibfield  {author} {\bibinfo {author} {\bibfnamefont {R.~N.}\ \bibnamefont
  {Silver}}, \bibinfo {author} {\bibfnamefont {D.~S.}\ \bibnamefont {Sivia}}, \
  and\ \bibinfo {author} {\bibfnamefont {J.~E.}\ \bibnamefont {Gubernatis}},\
  }\href {\doibase 10.1103/PhysRevB.41.2380} {\bibfield  {journal} {\bibinfo
  {journal} {Phys. Rev. B}\ }\textbf {\bibinfo {volume} {41}},\ \bibinfo
  {pages} {2380} (\bibinfo {year} {1990})}\BibitemShut {NoStop}%
\bibitem [{\citenamefont {Jarrell}\ and\ \citenamefont
  {Gubernatis}(1996)}]{Jarrell1996133}%
  \BibitemOpen
  \bibfield  {author} {\bibinfo {author} {\bibfnamefont {M.}~\bibnamefont
  {Jarrell}}\ and\ \bibinfo {author} {\bibfnamefont {J.}~\bibnamefont
  {Gubernatis}},\ }\href {\doibase 10.1016/0370-1573(95)00074-7} {\bibfield
  {journal} {\bibinfo  {journal} {Physics Reports}\ }\textbf {\bibinfo {volume}
  {269}},\ \bibinfo {pages} {133 } (\bibinfo {year} {1996})}\BibitemShut
  {NoStop}%
\bibitem [{\citenamefont {Bryan}(1990)}]{bryans}%
  \BibitemOpen
  \bibfield  {author} {\bibinfo {author} {\bibfnamefont {R.}~\bibnamefont
  {Bryan}},\ }\href {\doibase 10.1007/BF02427376} {\bibfield  {journal}
  {\bibinfo  {journal} {European Biophysics Journal}\ }\textbf {\bibinfo
  {volume} {18}},\ \bibinfo {pages} {165} (\bibinfo {year} {1990})}\BibitemShut
  {NoStop}%
\bibitem [{\citenamefont {Gubernatis}\ \emph {et~al.}(1991)\citenamefont
  {Gubernatis}, \citenamefont {Jarrell}, \citenamefont {Silver},\ and\
  \citenamefont {Sivia}}]{PhysRevB.44.6011}%
  \BibitemOpen
  \bibfield  {author} {\bibinfo {author} {\bibfnamefont {J.~E.}\ \bibnamefont
  {Gubernatis}}, \bibinfo {author} {\bibfnamefont {M.}~\bibnamefont {Jarrell}},
  \bibinfo {author} {\bibfnamefont {R.~N.}\ \bibnamefont {Silver}}, \ and\
  \bibinfo {author} {\bibfnamefont {D.~S.}\ \bibnamefont {Sivia}},\ }\href
  {\doibase 10.1103/PhysRevB.44.6011} {\bibfield  {journal} {\bibinfo
  {journal} {Phys. Rev. B}\ }\textbf {\bibinfo {volume} {44}},\ \bibinfo
  {pages} {6011} (\bibinfo {year} {1991})}\BibitemShut {NoStop}%
\bibitem [{\citenamefont {Fuchs}\ \emph {et~al.}(2010)\citenamefont {Fuchs},
  \citenamefont {Pruschke},\ and\ \citenamefont
  {Jarrell}}]{PhysRevE.81.056701}%
  \BibitemOpen
  \bibfield  {author} {\bibinfo {author} {\bibfnamefont {S.}~\bibnamefont
  {Fuchs}}, \bibinfo {author} {\bibfnamefont {T.}~\bibnamefont {Pruschke}}, \
  and\ \bibinfo {author} {\bibfnamefont {M.}~\bibnamefont {Jarrell}},\ }\href
  {\doibase 10.1103/PhysRevE.81.056701} {\bibfield  {journal} {\bibinfo
  {journal} {Phys. Rev. E}\ }\textbf {\bibinfo {volume} {81}},\ \bibinfo
  {pages} {056701} (\bibinfo {year} {2010})}\BibitemShut {NoStop}%
\bibitem [{\citenamefont {Dirks}\ \emph {et~al.}(2010)\citenamefont {Dirks},
  \citenamefont {Werner}, \citenamefont {Jarrell},\ and\ \citenamefont
  {Pruschke}}]{PhysRevE.82.026701}%
  \BibitemOpen
  \bibfield  {author} {\bibinfo {author} {\bibfnamefont {A.}~\bibnamefont
  {Dirks}}, \bibinfo {author} {\bibfnamefont {P.}~\bibnamefont {Werner}},
  \bibinfo {author} {\bibfnamefont {M.}~\bibnamefont {Jarrell}}, \ and\
  \bibinfo {author} {\bibfnamefont {T.}~\bibnamefont {Pruschke}},\ }\href
  {\doibase 10.1103/PhysRevE.82.026701} {\bibfield  {journal} {\bibinfo
  {journal} {Phys. Rev. E}\ }\textbf {\bibinfo {volume} {82}},\ \bibinfo
  {pages} {026701} (\bibinfo {year} {2010})}\BibitemShut {NoStop}%
\bibitem [{\citenamefont {Gunnarsson}\ \emph
  {et~al.}(2010{\natexlab{a}})\citenamefont {Gunnarsson}, \citenamefont
  {Haverkort},\ and\ \citenamefont {Sangiovanni}}]{PhysRevB.81.155107}%
  \BibitemOpen
  \bibfield  {author} {\bibinfo {author} {\bibfnamefont {O.}~\bibnamefont
  {Gunnarsson}}, \bibinfo {author} {\bibfnamefont {M.~W.}\ \bibnamefont
  {Haverkort}}, \ and\ \bibinfo {author} {\bibfnamefont {G.}~\bibnamefont
  {Sangiovanni}},\ }\href {\doibase 10.1103/PhysRevB.81.155107} {\bibfield
  {journal} {\bibinfo  {journal} {Phys. Rev. B}\ }\textbf {\bibinfo {volume}
  {81}},\ \bibinfo {pages} {155107} (\bibinfo {year}
  {2010}{\natexlab{a}})}\BibitemShut {NoStop}%
\bibitem [{\citenamefont {Gunnarsson}\ \emph
  {et~al.}(2010{\natexlab{b}})\citenamefont {Gunnarsson}, \citenamefont
  {Haverkort},\ and\ \citenamefont {Sangiovanni}}]{PhysRevB.82.165125}%
  \BibitemOpen
  \bibfield  {author} {\bibinfo {author} {\bibfnamefont {O.}~\bibnamefont
  {Gunnarsson}}, \bibinfo {author} {\bibfnamefont {M.~W.}\ \bibnamefont
  {Haverkort}}, \ and\ \bibinfo {author} {\bibfnamefont {G.}~\bibnamefont
  {Sangiovanni}},\ }\href {\doibase 10.1103/PhysRevB.82.165125} {\bibfield
  {journal} {\bibinfo  {journal} {Phys. Rev. B}\ }\textbf {\bibinfo {volume}
  {82}},\ \bibinfo {pages} {165125} (\bibinfo {year}
  {2010}{\natexlab{b}})}\BibitemShut {NoStop}%
\bibitem [{\citenamefont {Creffield}\ \emph {et~al.}(1995)\citenamefont
  {Creffield}, \citenamefont {Klepfish}, \citenamefont {Pike},\ and\
  \citenamefont {Sarkar}}]{PhysRevLett.75.517}%
  \BibitemOpen
  \bibfield  {author} {\bibinfo {author} {\bibfnamefont {C.~E.}\ \bibnamefont
  {Creffield}}, \bibinfo {author} {\bibfnamefont {E.~G.}\ \bibnamefont
  {Klepfish}}, \bibinfo {author} {\bibfnamefont {E.~R.}\ \bibnamefont {Pike}},
  \ and\ \bibinfo {author} {\bibfnamefont {S.}~\bibnamefont {Sarkar}},\ }\href
  {\doibase 10.1103/PhysRevLett.75.517} {\bibfield  {journal} {\bibinfo
  {journal} {Phys. Rev. Lett.}\ }\textbf {\bibinfo {volume} {75}},\ \bibinfo
  {pages} {517} (\bibinfo {year} {1995})}\BibitemShut {NoStop}%
\bibitem [{\citenamefont {A.~Tikhonov}\ and\ \citenamefont
  {Yagola}(1995)}]{Tikhonov}%
  \BibitemOpen
  \bibfield  {author} {\bibinfo {author} {\bibfnamefont {V.~S.}\ \bibnamefont
  {A.~Tikhonov}, \bibfnamefont {A.~Goncharsky}}\ and\ \bibinfo {author}
  {\bibfnamefont {A.}~\bibnamefont {Yagola}},\ }\href@noop {} {\emph {\bibinfo
  {title} {Numerical Methods for the Solution of Ill-Posed Problems}}}\
  (\bibinfo  {publisher} {Springer-Science+Business Media, B.V.},\ \bibinfo
  {address} {Moscow},\ \bibinfo {year} {1995})\BibitemShut {NoStop}%
\bibitem [{\citenamefont {Lawson}\ and\ \citenamefont {Hanson}(1995)}]{NNLS}%
  \BibitemOpen
  \bibfield  {author} {\bibinfo {author} {\bibfnamefont {C.~L.}\ \bibnamefont
  {Lawson}}\ and\ \bibinfo {author} {\bibfnamefont {R.~J.}\ \bibnamefont
  {Hanson}},\ }\href@noop {} {\emph {\bibinfo {title} {Solving Least Squares
  Problems}}}\ (\bibinfo  {publisher} {Society for Industrial and Applied
  Mathematics Philadelphia},\ \bibinfo {address} {Philadelphia},\ \bibinfo
  {year} {1995})\BibitemShut {NoStop}%
\bibitem [{\citenamefont {Prokof’ev}\ and\ \citenamefont
  {Svistunov}(2013)}]{MCC}%
  \BibitemOpen
  \bibfield  {author} {\bibinfo {author} {\bibfnamefont {N.}~\bibnamefont
  {Prokof’ev}}\ and\ \bibinfo {author} {\bibfnamefont {B.}~\bibnamefont
  {Svistunov}},\ }\href {\doibase 10.1134/S002136401311009X} {\bibfield
  {journal} {\bibinfo  {journal} {JETP Letters}\ }\textbf {\bibinfo {volume}
  {97}},\ \bibinfo {pages} {649} (\bibinfo {year} {2013})}\BibitemShut
  {NoStop}%
\bibitem [{\citenamefont {{Krivenko}}\ and\ \citenamefont
  {{Rubtsov}}(2006)}]{2006cond.mat.12233K}%
  \BibitemOpen
  \bibfield  {author} {\bibinfo {author} {\bibfnamefont {I.~S.}\ \bibnamefont
  {{Krivenko}}}\ and\ \bibinfo {author} {\bibfnamefont {A.~N.}\ \bibnamefont
  {{Rubtsov}}},\ }\href@noop {} {\bibfield  {journal} {\bibinfo  {journal}
  {arXiv:cond-mat/0612233 (unpublished)}\  } } 
  \BibitemShut
  {NoStop}%
\bibitem [{\citenamefont {Vafayi}\ and\ \citenamefont
  {Gunnarsson}(2007)}]{PhysRevB.76.035115}%
  \BibitemOpen
  \bibfield  {author} {\bibinfo {author} {\bibfnamefont {K.}~\bibnamefont
  {Vafayi}}\ and\ \bibinfo {author} {\bibfnamefont {O.}~\bibnamefont
  {Gunnarsson}},\ }\href {\doibase 10.1103/PhysRevB.76.035115} {\bibfield
  {journal} {\bibinfo  {journal} {Phys. Rev. B}\ }\textbf {\bibinfo {volume}
  {76}},\ \bibinfo {pages} {035115} (\bibinfo {year} {2007})}\BibitemShut
  {NoStop}%
\bibitem [{\citenamefont {Sandvik}(1998)}]{PhysRevB.57.10287}%
  \BibitemOpen
  \bibfield  {author} {\bibinfo {author} {\bibfnamefont {A.~W.}\ \bibnamefont
  {Sandvik}},\ }\href {\doibase 10.1103/PhysRevB.57.10287} {\bibfield
  {journal} {\bibinfo  {journal} {Phys. Rev. B}\ }\textbf {\bibinfo {volume}
  {57}},\ \bibinfo {pages} {10287} (\bibinfo {year} {1998})}\BibitemShut
  {NoStop}%
\bibitem [{\citenamefont {Mishchenko}\ \emph {et~al.}(2000)\citenamefont
  {Mishchenko}, \citenamefont {Prokof'ev}, \citenamefont {Sakamoto},\ and\
  \citenamefont {Svistunov}}]{PhysRevB.62.6317}%
  \BibitemOpen
  \bibfield  {author} {\bibinfo {author} {\bibfnamefont {A.~S.}\ \bibnamefont
  {Mishchenko}}, \bibinfo {author} {\bibfnamefont {N.~V.}\ \bibnamefont
  {Prokof'ev}}, \bibinfo {author} {\bibfnamefont {A.}~\bibnamefont {Sakamoto}},
  \ and\ \bibinfo {author} {\bibfnamefont {B.~V.}\ \bibnamefont {Svistunov}},\
  }\href {\doibase 10.1103/PhysRevB.62.6317} {\bibfield  {journal} {\bibinfo
  {journal} {Phys. Rev. B}\ }\textbf {\bibinfo {volume} {62}},\ \bibinfo
  {pages} {6317} (\bibinfo {year} {2000})}\BibitemShut {NoStop}%
\bibitem [{\citenamefont {Staar}\ \emph {et~al.}(2014)\citenamefont {Staar},
  \citenamefont {Ydens}, \citenamefont {Kozhevnikov}, \citenamefont {Locquet},\
  and\ \citenamefont {Schulthess}}]{staar14}%
  \BibitemOpen
  \bibfield  {author} {\bibinfo {author} {\bibfnamefont {P.}~\bibnamefont
  {Staar}}, \bibinfo {author} {\bibfnamefont {B.}~\bibnamefont {Ydens}},
  \bibinfo {author} {\bibfnamefont {A.}~\bibnamefont {Kozhevnikov}}, \bibinfo
  {author} {\bibfnamefont {J.-P.}\ \bibnamefont {Locquet}}, \ and\ \bibinfo
  {author} {\bibfnamefont {T.}~\bibnamefont {Schulthess}},\ }\href {\doibase
  10.1103/PhysRevB.89.245114} {\bibfield  {journal} {\bibinfo  {journal} {Phys.
  Rev. B}\ }\textbf {\bibinfo {volume} {89}},\ \bibinfo {pages} {245114}
  (\bibinfo {year} {2014})}\BibitemShut {NoStop}%
\bibitem [{\citenamefont {Rota}\ \emph {et~al.}(2015)\citenamefont {Rota},
  \citenamefont {Casulleras}, \citenamefont {Mazzanti},\ and\ \citenamefont
  {Boronat}}]{Little_wicks}%
  \BibitemOpen
  \bibfield  {author} {\bibinfo {author} {\bibfnamefont {R.}~\bibnamefont
  {Rota}}, \bibinfo {author} {\bibfnamefont {J.}~\bibnamefont {Casulleras}},
  \bibinfo {author} {\bibfnamefont {F.}~\bibnamefont {Mazzanti}}, \ and\
  \bibinfo {author} {\bibfnamefont {J.}~\bibnamefont {Boronat}},\ }\href
  {\doibase http://dx.doi.org/10.1063/1.4914995} {\bibfield  {journal}
  {\bibinfo  {journal} {The Journal of Chemical Physics}\ }\textbf {\bibinfo
  {volume} {142}},\ \bibinfo {eid} {114114} (\bibinfo {year} {2015}),\
  http://dx.doi.org/10.1063/1.4914995}\BibitemShut {NoStop}%
\bibitem [{\citenamefont {Baker}(1975)}]{Baker_with_Thiele}%
  \BibitemOpen
  \bibfield  {author} {\bibinfo {author} {\bibfnamefont {G.~A.}\ \bibnamefont
  {Baker}},\ }\href@noop {} {\emph {\bibinfo {title} {Essentials of Pad\'{e}
  Approximants}}}\ (\bibinfo  {publisher} {Academic Press},\ \bibinfo {address}
  {New York},\ \bibinfo {year} {1975})\BibitemShut {NoStop}%
\bibitem [{\citenamefont {Vidberg}\ and\ \citenamefont
  {Serene}(1977)}]{vidberg77jltp29:179}%
  \BibitemOpen
  \bibfield  {author} {\bibinfo {author} {\bibfnamefont {H.~J.}\ \bibnamefont
  {Vidberg}}\ and\ \bibinfo {author} {\bibfnamefont {J.~W.}\ \bibnamefont
  {Serene}},\ }\href@noop {} {\bibfield  {journal} {\bibinfo  {journal} {J. Low
  Temp. Phys.}\ }\textbf {\bibinfo {volume} {29}},\ \bibinfo {pages} {179}
  (\bibinfo {year} {1977})}\BibitemShut {NoStop}%
\bibitem [{\citenamefont {Beach}\ \emph {et~al.}(2000)\citenamefont {Beach},
  \citenamefont {Gooding},\ and\ \citenamefont {Marsiglio}}]{beach}%
  \BibitemOpen
  \bibfield  {author} {\bibinfo {author} {\bibfnamefont {K.~S.~D.}\
  \bibnamefont {Beach}}, \bibinfo {author} {\bibfnamefont {R.~J.}\ \bibnamefont
  {Gooding}}, \ and\ \bibinfo {author} {\bibfnamefont {F.}~\bibnamefont
  {Marsiglio}},\ }\href {\doibase 10.1103/PhysRevB.61.5147} {\bibfield
  {journal} {\bibinfo  {journal} {Phys. Rev. B}\ }\textbf {\bibinfo {volume}
  {61}},\ \bibinfo {pages} {5147} (\bibinfo {year} {2000})}\BibitemShut
  {NoStop}%
\bibitem [{\citenamefont {Gull}\ \emph {et~al.}(2011)\citenamefont {Gull},
  \citenamefont {Millis}, \citenamefont {Lichtenstein}, \citenamefont
  {Rubtsov}, \citenamefont {Troyer},\ and\ \citenamefont {Werner}}]{gull11}%
  \BibitemOpen
  \bibfield  {author} {\bibinfo {author} {\bibfnamefont {E.}~\bibnamefont
  {Gull}}, \bibinfo {author} {\bibfnamefont {A.~J.}\ \bibnamefont {Millis}},
  \bibinfo {author} {\bibfnamefont {A.~I.}\ \bibnamefont {Lichtenstein}},
  \bibinfo {author} {\bibfnamefont {A.~N.}\ \bibnamefont {Rubtsov}}, \bibinfo
  {author} {\bibfnamefont {M.}~\bibnamefont {Troyer}}, \ and\ \bibinfo {author}
  {\bibfnamefont {P.}~\bibnamefont {Werner}},\ }\href {\doibase
  10.1103/RevModPhys.83.349} {\bibfield  {journal} {\bibinfo  {journal} {Rev.
  Mod. Phys.}\ }\textbf {\bibinfo {volume} {83}},\ \bibinfo {pages} {349}
  (\bibinfo {year} {2011})}\BibitemShut {NoStop}%
\bibitem [{\citenamefont {\"Ostlin}\ \emph {et~al.}(2012)\citenamefont
  {\"Ostlin}, \citenamefont {Chioncel},\ and\ \citenamefont {Vitos}}]{Andreas}%
  \BibitemOpen
  \bibfield  {author} {\bibinfo {author} {\bibfnamefont {A.}~\bibnamefont
  {\"Ostlin}}, \bibinfo {author} {\bibfnamefont {L.}~\bibnamefont {Chioncel}},
  \ and\ \bibinfo {author} {\bibfnamefont {L.}~\bibnamefont {Vitos}},\ }\href
  {\doibase 10.1103/PhysRevB.86.235107} {\bibfield  {journal} {\bibinfo
  {journal} {Phys. Rev. B}\ }\textbf {\bibinfo {volume} {86}},\ \bibinfo
  {pages} {235107} (\bibinfo {year} {2012})}\BibitemShut {NoStop}%
\bibitem [{\citenamefont {{Sokolovski}}\ \emph {et~al.}(2011)\citenamefont
  {{Sokolovski}}, \citenamefont {{Akhmatskaya}},\ and\ \citenamefont
  {{Sen}}}]{2011CoPhC.182..448S}%
  \BibitemOpen
  \bibfield  {author} {\bibinfo {author} {\bibfnamefont {D.}~\bibnamefont
  {{Sokolovski}}}, \bibinfo {author} {\bibfnamefont {E.}~\bibnamefont
  {{Akhmatskaya}}}, \ and\ \bibinfo {author} {\bibfnamefont {S.~K.}\
  \bibnamefont {{Sen}}},\ }\href {\doibase 10.1016/j.cpc.2010.10.002}
  {\bibfield  {journal} {\bibinfo  {journal} {Computer Physics Communications}\
  }\textbf {\bibinfo {volume} {182}},\ \bibinfo {pages} {448} (\bibinfo {year}
  {2011})}\BibitemShut {NoStop}%
\bibitem [{\citenamefont {\ifmmode \check{Z}\else~\v{Z}\fi{}. Osolin}\ and\
  \citenamefont {\ifmmode~\check{Z}\else
  \v{Z}\fi{}itko}(2013)}]{PhysRevB.87.245135}%
  \BibitemOpen
  \bibfield  {author} {\bibinfo {author} {\bibnamefont {\ifmmode
  \check{Z}\else~\v{Z}\fi{}. Osolin}}\ and\ \bibinfo {author} {\bibfnamefont
  {R.}~\bibnamefont {\ifmmode~\check{Z}\else \v{Z}\fi{}itko}},\ }\href
  {\doibase 10.1103/PhysRevB.87.245135} {\bibfield  {journal} {\bibinfo
  {journal} {Phys. Rev. B}\ }\textbf {\bibinfo {volume} {87}},\ \bibinfo
  {pages} {245135} (\bibinfo {year} {2013})}\BibitemShut {NoStop}%
\bibitem [{\citenamefont {Grechnev}\ \emph {et~al.}(2007)\citenamefont
  {Grechnev}, \citenamefont {Di~Marco}, \citenamefont {Katsnelson},
  \citenamefont {Lichtenstein}, \citenamefont {Wills},\ and\ \citenamefont
  {Eriksson}}]{grechnev07prb76:035107}%
  \BibitemOpen
  \bibfield  {author} {\bibinfo {author} {\bibfnamefont {A.}~\bibnamefont
  {Grechnev}}, \bibinfo {author} {\bibfnamefont {I.}~\bibnamefont {Di~Marco}},
  \bibinfo {author} {\bibfnamefont {M.~I.}\ \bibnamefont {Katsnelson}},
  \bibinfo {author} {\bibfnamefont {A.~I.}\ \bibnamefont {Lichtenstein}},
  \bibinfo {author} {\bibfnamefont {J.}~\bibnamefont {Wills}}, \ and\ \bibinfo
  {author} {\bibfnamefont {O.}~\bibnamefont {Eriksson}},\ }\href {\doibase
  10.1103/PhysRevB.76.035107} {\bibfield  {journal} {\bibinfo  {journal} {Phys.
  Rev. B}\ }\textbf {\bibinfo {volume} {76}},\ \bibinfo {pages} {035107}
  (\bibinfo {year} {2007})}\BibitemShut {NoStop}%
\bibitem [{\citenamefont {Wang}\ \emph {et~al.}(2009)\citenamefont {Wang},
  \citenamefont {Gull}, \citenamefont {de' Medici}, \citenamefont {Capone},\
  and\ \citenamefont {Millis}}]{PhysRevB.80.045101}%
  \BibitemOpen
  \bibfield  {author} {\bibinfo {author} {\bibfnamefont {X.}~\bibnamefont
  {Wang}}, \bibinfo {author} {\bibfnamefont {E.}~\bibnamefont {Gull}}, \bibinfo
  {author} {\bibfnamefont {L.}~\bibnamefont {de' Medici}}, \bibinfo {author}
  {\bibfnamefont {M.}~\bibnamefont {Capone}}, \ and\ \bibinfo {author}
  {\bibfnamefont {A.~J.}\ \bibnamefont {Millis}},\ }\href {\doibase
  10.1103/PhysRevB.80.045101} {\bibfield  {journal} {\bibinfo  {journal} {Phys.
  Rev. B}\ }\textbf {\bibinfo {volume} {80}},\ \bibinfo {pages} {045101}
  (\bibinfo {year} {2009})}\BibitemShut {NoStop}%
\bibitem [{\citenamefont {Wills}\ \emph {et~al.}(2000)\citenamefont {Wills},
  \citenamefont {Delin},\ and\ \citenamefont {Eriksson}}]{RSPtBook}%
  \BibitemOpen
  \bibfield  {author} {\bibinfo {author} {\bibfnamefont {J.}~\bibnamefont
  {Wills}}, \bibinfo {author} {\bibfnamefont {A.}~\bibnamefont {Delin}}, \ and\
  \bibinfo {author} {\bibfnamefont {O.}~\bibnamefont {Eriksson}},\ }\href@noop
  {} {\emph {\bibinfo {title} {Full-Potential Electronic Structure Method}}}\
  (\bibinfo  {publisher} {Springer},\ \bibinfo {address} {New York},\ \bibinfo
  {year} {2000})\BibitemShut {NoStop}%
\bibitem [{\citenamefont {Thunstr\"om}\ \emph {et~al.}(2009)\citenamefont
  {Thunstr\"om}, \citenamefont {Di~Marco}, \citenamefont {Grechnev},
  \citenamefont {Leb\`egue}, \citenamefont {Katsnelson}, \citenamefont
  {Svane},\ and\ \citenamefont {Eriksson}}]{PhysRevB.79.165104}%
  \BibitemOpen
  \bibfield  {author} {\bibinfo {author} {\bibfnamefont {P.}~\bibnamefont
  {Thunstr\"om}}, \bibinfo {author} {\bibfnamefont {I.}~\bibnamefont
  {Di~Marco}}, \bibinfo {author} {\bibfnamefont {A.}~\bibnamefont {Grechnev}},
  \bibinfo {author} {\bibfnamefont {S.}~\bibnamefont {Leb\`egue}}, \bibinfo
  {author} {\bibfnamefont {M.~I.}\ \bibnamefont {Katsnelson}}, \bibinfo
  {author} {\bibfnamefont {A.}~\bibnamefont {Svane}}, \ and\ \bibinfo {author}
  {\bibfnamefont {O.}~\bibnamefont {Eriksson}},\ }\href {\doibase
  10.1103/PhysRevB.79.165104} {\bibfield  {journal} {\bibinfo  {journal} {Phys.
  Rev. B}\ }\textbf {\bibinfo {volume} {79}},\ \bibinfo {pages} {165104}
  (\bibinfo {year} {2009})}\BibitemShut {NoStop}%
\bibitem [{\citenamefont {Gr{\aa}n\"as}\ \emph {et~al.}(2012)\citenamefont
  {Gr{\aa}n\"as}, \citenamefont {{Di Marco}}, \citenamefont {Thunstr\"om},
  \citenamefont {Nordstr\"om}, \citenamefont {Eriksson}, \citenamefont
  {Bj\"orkman},\ and\ \citenamefont {Wills}}]{granas12}%
  \BibitemOpen
  \bibfield  {author} {\bibinfo {author} {\bibfnamefont {O.}~\bibnamefont
  {Gr{\aa}n\"as}}, \bibinfo {author} {\bibfnamefont {I.}~\bibnamefont {{Di
  Marco}}}, \bibinfo {author} {\bibfnamefont {P.}~\bibnamefont {Thunstr\"om}},
  \bibinfo {author} {\bibfnamefont {L.}~\bibnamefont {Nordstr\"om}}, \bibinfo
  {author} {\bibfnamefont {O.}~\bibnamefont {Eriksson}}, \bibinfo {author}
  {\bibfnamefont {T.}~\bibnamefont {Bj\"orkman}}, \ and\ \bibinfo {author}
  {\bibfnamefont {J.}~\bibnamefont {Wills}},\ }\href {\doibase
  10.1016/j.commatsci.2011.11.032} {\bibfield  {journal} {\bibinfo  {journal}
  {Computational Materials Science}\ }\textbf {\bibinfo {volume} {55}},\
  \bibinfo {pages} {295 } (\bibinfo {year} {2012})}\BibitemShut {NoStop}%
\bibitem [{\citenamefont {Nakata}()}]{mpack}%
  \BibitemOpen
  \bibfield  {author} {\bibinfo {author} {\bibfnamefont {M.}~\bibnamefont
  {Nakata}},\ }\href@noop {} {\emph {\bibinfo {title} {The MPACK
  (MBLAS/MLAPACK): A Multiple Precision Arithmetic Version of BLAS and
  LAPACK}}},\ \bibinfo {address} {http://mplapack.sourceforge.net},\ \bibinfo
  {edition} {0th}\ ed.\BibitemShut {Stop}%
\bibitem [{\citenamefont {Anderson}\ \emph {et~al.}(1999)\citenamefont
  {Anderson}, \citenamefont {Bai}, \citenamefont {Bischof}, \citenamefont
  {Blackford}, \citenamefont {Demmel}, \citenamefont {Dongarra}, \citenamefont
  {Du~Croz}, \citenamefont {Greenbaum}, \citenamefont {Hammarling},
  \citenamefont {McKenney},\ and\ \citenamefont {Sorensen}}]{laug}%
  \BibitemOpen
  \bibfield  {author} {\bibinfo {author} {\bibfnamefont {E.}~\bibnamefont
  {Anderson}}, \bibinfo {author} {\bibfnamefont {Z.}~\bibnamefont {Bai}},
  \bibinfo {author} {\bibfnamefont {C.}~\bibnamefont {Bischof}}, \bibinfo
  {author} {\bibfnamefont {S.}~\bibnamefont {Blackford}}, \bibinfo {author}
  {\bibfnamefont {J.}~\bibnamefont {Demmel}}, \bibinfo {author} {\bibfnamefont
  {J.}~\bibnamefont {Dongarra}}, \bibinfo {author} {\bibfnamefont
  {J.}~\bibnamefont {Du~Croz}}, \bibinfo {author} {\bibfnamefont
  {A.}~\bibnamefont {Greenbaum}}, \bibinfo {author} {\bibfnamefont
  {S.}~\bibnamefont {Hammarling}}, \bibinfo {author} {\bibfnamefont
  {A.}~\bibnamefont {McKenney}}, \ and\ \bibinfo {author} {\bibfnamefont
  {D.}~\bibnamefont {Sorensen}},\ }\href@noop {} {\emph {\bibinfo {title}
  {{LAPACK} Users' Guide}}},\ \bibinfo {edition} {3rd}\ ed.\ (\bibinfo
  {publisher} {Society for Industrial and Applied Mathematics},\ \bibinfo
  {address} {Philadelphia, PA},\ \bibinfo {year} {1999})\BibitemShut {NoStop}%
\bibitem [{\citenamefont {Haldane}(1988)}]{PhysRevLett.61.2015}%
  \BibitemOpen
  \bibfield  {author} {\bibinfo {author} {\bibfnamefont {F.~D.~M.}\
  \bibnamefont {Haldane}},\ }\href {\doibase 10.1103/PhysRevLett.61.2015}
  {\bibfield  {journal} {\bibinfo  {journal} {Phys. Rev. Lett.}\ }\textbf
  {\bibinfo {volume} {61}},\ \bibinfo {pages} {2015} (\bibinfo {year}
  {1988})}\BibitemShut {NoStop}%
\bibitem [{\citenamefont {Thunstr\"om}\ \emph {et~al.}(2012)\citenamefont
  {Thunstr\"om}, \citenamefont {Di~Marco},\ and\ \citenamefont
  {Eriksson}}]{thunstrom12prl109:186401}%
  \BibitemOpen
  \bibfield  {author} {\bibinfo {author} {\bibfnamefont {P.}~\bibnamefont
  {Thunstr\"om}}, \bibinfo {author} {\bibfnamefont {I.}~\bibnamefont
  {Di~Marco}}, \ and\ \bibinfo {author} {\bibfnamefont {O.}~\bibnamefont
  {Eriksson}},\ }\href {\doibase 10.1103/PhysRevLett.109.186401} {\bibfield
  {journal} {\bibinfo  {journal} {Phys. Rev. Lett.}\ }\textbf {\bibinfo
  {volume} {109}},\ \bibinfo {pages} {186401} (\bibinfo {year}
  {2012})}\BibitemShut {NoStop}%
\bibitem [{\citenamefont {Lu}\ \emph {et~al.}(2014)\citenamefont {Lu},
  \citenamefont {H\"oppner}, \citenamefont {Gunnarsson},\ and\ \citenamefont
  {Haverkort}}]{lu14prb90_085102}%
  \BibitemOpen
  \bibfield  {author} {\bibinfo {author} {\bibfnamefont {Y.}~\bibnamefont
  {Lu}}, \bibinfo {author} {\bibfnamefont {M.}~\bibnamefont {H\"oppner}},
  \bibinfo {author} {\bibfnamefont {O.}~\bibnamefont {Gunnarsson}}, \ and\
  \bibinfo {author} {\bibfnamefont {M.~W.}\ \bibnamefont {Haverkort}},\ }\href
  {\doibase 10.1103/PhysRevB.90.085102} {\bibfield  {journal} {\bibinfo
  {journal} {Phys. Rev. B}\ }\textbf {\bibinfo {volume} {90}},\ \bibinfo
  {pages} {085102} (\bibinfo {year} {2014})}\BibitemShut {NoStop}%
\bibitem [{\citenamefont {Koloren\ifmmode~\check{c}\else \v{c}\fi{}}\ \emph
  {et~al.}(2012)\citenamefont {Koloren\ifmmode~\check{c}\else \v{c}\fi{}},
  \citenamefont {Poteryaev},\ and\ \citenamefont
  {Lichtenstein}}]{kolorenc12prb85_235136}%
  \BibitemOpen
  \bibfield  {author} {\bibinfo {author} {\bibfnamefont {J.}~\bibnamefont
  {Koloren\ifmmode~\check{c}\else \v{c}\fi{}}}, \bibinfo {author}
  {\bibfnamefont {A.~I.}\ \bibnamefont {Poteryaev}}, \ and\ \bibinfo {author}
  {\bibfnamefont {A.~I.}\ \bibnamefont {Lichtenstein}},\ }\href {\doibase
  10.1103/PhysRevB.85.235136} {\bibfield  {journal} {\bibinfo  {journal} {Phys.
  Rev. B}\ }\textbf {\bibinfo {volume} {85}},\ \bibinfo {pages} {235136}
  (\bibinfo {year} {2012})}\BibitemShut {NoStop}%
\bibitem [{\citenamefont {Granath}\ and\ \citenamefont
  {Strand}(2012)}]{granath12prb86:115111}%
  \BibitemOpen
  \bibfield  {author} {\bibinfo {author} {\bibfnamefont {M.}~\bibnamefont
  {Granath}}\ and\ \bibinfo {author} {\bibfnamefont {H.~U.~R.}\ \bibnamefont
  {Strand}},\ }\href {\doibase 10.1103/PhysRevB.86.115111} {\bibfield
  {journal} {\bibinfo  {journal} {Phys. Rev. B}\ }\textbf {\bibinfo {volume}
  {86}},\ \bibinfo {pages} {115111} (\bibinfo {year} {2012})}\BibitemShut
  {NoStop}%
\bibitem [{\citenamefont {Chioncel}\ \emph {et~al.}(2003)\citenamefont
  {Chioncel}, \citenamefont {Vitos}, \citenamefont {Abrikosov}, \citenamefont
  {Koll\'ar}, \citenamefont {Katsnelson},\ and\ \citenamefont
  {Lichtenstein}}]{PhysRevB.67.235106}%
  \BibitemOpen
  \bibfield  {author} {\bibinfo {author} {\bibfnamefont {L.}~\bibnamefont
  {Chioncel}}, \bibinfo {author} {\bibfnamefont {L.}~\bibnamefont {Vitos}},
  \bibinfo {author} {\bibfnamefont {I.~A.}\ \bibnamefont {Abrikosov}}, \bibinfo
  {author} {\bibfnamefont {J.}~\bibnamefont {Koll\'ar}}, \bibinfo {author}
  {\bibfnamefont {M.~I.}\ \bibnamefont {Katsnelson}}, \ and\ \bibinfo {author}
  {\bibfnamefont {A.~I.}\ \bibnamefont {Lichtenstein}},\ }\href {\doibase
  10.1103/PhysRevB.67.235106} {\bibfield  {journal} {\bibinfo  {journal} {Phys.
  Rev. B}\ }\textbf {\bibinfo {volume} {67}},\ \bibinfo {pages} {235106}
  (\bibinfo {year} {2003})}\BibitemShut {NoStop}%
\bibitem [{\citenamefont {Min\'{a}r}\ \emph {et~al.}(2005)\citenamefont
  {Min\'{a}r}, \citenamefont {Chioncel}, \citenamefont {Perlov}, \citenamefont
  {Ebert}, \citenamefont {Katsnelson},\ and\ \citenamefont
  {Lichtenstein}}]{minar05prb72:045125}%
  \BibitemOpen
  \bibfield  {author} {\bibinfo {author} {\bibfnamefont {J.}~\bibnamefont
  {Min\'{a}r}}, \bibinfo {author} {\bibfnamefont {L.}~\bibnamefont {Chioncel}},
  \bibinfo {author} {\bibfnamefont {A.}~\bibnamefont {Perlov}}, \bibinfo
  {author} {\bibfnamefont {H.}~\bibnamefont {Ebert}}, \bibinfo {author}
  {\bibfnamefont {M.~I.}\ \bibnamefont {Katsnelson}}, \ and\ \bibinfo {author}
  {\bibfnamefont {A.~I.}\ \bibnamefont {Lichtenstein}},\ }\href@noop {}
  {\bibfield  {journal} {\bibinfo  {journal} {Phys. Rev. B}\ }\textbf {\bibinfo
  {volume} {72}},\ \bibinfo {pages} {045125} (\bibinfo {year}
  {2005})}\BibitemShut {NoStop}%
\bibitem [{\citenamefont {Hansen}(2000)}]{L-curve}%
  \BibitemOpen
  \bibfield  {author} {\bibinfo {author} {\bibfnamefont {P.~C.}\ \bibnamefont
  {Hansen}},\ }in\ \href@noop {} {\emph {\bibinfo {booktitle} {in Computational
  Inverse Problems in Electrocardiology, ed. P. Johnston, Advances in
  Computational Bioengineering}}}\ (\bibinfo  {publisher} {WIT Press},\
  \bibinfo {address} {Southampton},\ \bibinfo {year} {2000})\ pp.\ \bibinfo {pages} {119--142}\BibitemShut
  {NoStop}%
\end{thebibliography}

\bibliographystyle{apsrev4-1}
\end{document}